\documentclass[reprint,groupedaddress,amsmath,amssymb,pra]{revtex4-2}

\usepackage{hyperref}
\usepackage{amsmath} 
\usepackage{amssymb}
\usepackage[pdftex]{graphicx}
\usepackage{microtype}
\usepackage{cancel}
\usepackage{mathrsfs}

\usepackage{setspace}
\usepackage{titlesec}

\usepackage{color}
\usepackage[shortlabels]{enumitem}
\usepackage[normalem]{ulem}
\usepackage{booktabs}

\newcommand{\mrm}{\mathrm}

\newcommand{\ket}[1]
    {{\left| {#1} \right\rangle}}
\newcommand{\bra}[1]
    {{\left\langle {#1} \right|}}

\newcommand{\braket}[2]
    {{\left\langle {#1}|{#2} \right\rangle}}
\newcommand{\avg}[1]
    {\left\langle {#1} 
    \right\rangle }

\newcommand{\Evec}
    {\vec{\mathcal{E}}}

\newcommand{\Bvec}
    {\vec{\mathcal{B}}}

\newcommand{\Esca}{\mathcal{E}}

\newcommand{\Bsca}{\mathcal{B}}

\newcommand{\eu}{$^{153}$Eu}

\begin{document}

\title{Nuclear T-violation search using octupole-deformed nuclei in a crystal}

\author{H. D. Ramachandran}
\author{A. C. Vutha}
\affiliation{Department of Physics, University of Toronto, Toronto ON M5S 1A7, Canada}

\begin{abstract}
Precision measurements with atoms and molecules can search for subtle violations of time-reversal symmetry (T) in nuclei, and thereby probe a variety of new physics models. We present a detailed scheme for a nuclear T-violation search experiment using $^{153}$Eu$^{3+}$ ions doped in non-centrosymmetric sites within a Y$_2$SiO$_5$ crystal. The ions in this solid contain nuclei that are highly sensitive to T-violation, and avail of large atomic enhancements by being polarized within the solid. But in particular, the system and methods that we discuss here enable the use of vast numbers of nuclei trapped in crystals, while also offering a number of stringent tests to ward off systematic errors. Our approach maps out a path to probe new physics at the PeV energy scale.
\end{abstract}

\maketitle

\section{Introduction}
New sources of time-reversal symmetry (T) violation are necessary to explain the observed excess of matter over antimatter in the universe \cite{Sakharov1991}. Theories beyond the Standard Model, and existing experimental constraints, indicate that the energy scale for this new physics is inaccessible to present-day particle colliders \cite{Engel2013, DeMille2017}. 

New physics sources of T-violation coupled to hadrons lead to P-odd, T-odd nuclear moments, which lead to measurable shifts in the energy levels of atoms \cite{Sushkov1984}. These symmetry-breaking nuclear moments (e.g., nuclear Schiff moments, magnetic quadrupole moments) can be enhanced by almost three orders of magnitude in nuclei that exhibit a combination of quadrupole and octupole deformations \cite{Skripnikov2020}. The energy shifts induced in atoms by these nuclear moments are magnified in atoms that are electrically polarized. Therefore T-violating energy shifts can be enhanced for deformed nuclei within polar molecules, wherein the constituent atomic ions are strongly polarized: this property has led to a number of experiments and proposals to study nuclear T-violation using polar molecules \cite{Sandars1967, Wilkening1984, Isaev2010, Grasdijk2021, DeMille2021}.

But polar molecules are not the only way to polarize atoms. Atomic ions in non-centrosymmetric sites in a crystal are also strongly electrically polarized. In effect, each ion in such a crystal can be considered as being part of a very large polar molecule. Therefore experiments with nuclei enclosed by polarized ions can avail of enhanced internal electric fields, just as in polar molecule experiments, while also gaining sensitivity due to the enormous quantity of trapped ions that can be interrogated in solid-state samples. The potential of experiments with such doped solids has been previously suggested by Singh \cite{Singh2019} and Flambaum \emph{et al.} \cite{Flambaum2020b}. 

However, there are two main challenges that need to be addressed before a T-violation search experiment in a solid can be practically realized. 
\begin{enumerate}[i)]
\item Although the ions in non-centrosymmetric sites are electrically polarized, the \emph{average} electrical polarization over the crystal can still be zero: in other words, the electrical polarization vectors of ions within the crystal can point in many possible directions. Therefore, experiments on the full ensemble of ions in the crystal cannot measure T-violating energy shifts, unless there is either a biased distribution of polarization vectors (e.g., in a ferroelectric crystal), or some way to separately measure signals from differently polarized ions. 

\item Optical and nuclear spin transitions of ions in solids undergo decoherence and inhomogeneous broadening in the solid-state, due to impurities, defects, phonons and the crystal fields from neighboring ions. Therefore, the techniques used for high signal-to-noise ratio measurements in gas-phase experiments, such as optical pumping and optically-detected magnetic resonance \cite{Graner2016, Grasdijk2021}, cannot be trivially ported over to solid-state experiments. 
\end{enumerate} 

Thus, we believe that new measurement schemes are required in order to make practical advances in new physics searches in solids. These measurement schemes must also provide protection from systematic errors so that the high statistical sensitivity of solid-state systems can be gainfully exploited. In this paper, we discuss the details of a measurement scheme that addresses both these challenges, expanding on the ideas presented in Ref.\ \cite{Ramachandran2022}. For definiteness, we focus this discussion on the Eu:YSO system: $^{153}$Eu$^{3+}$ ions doped into yttrium orthosilicate (Y$_2$SiO$_5$), although the methods described here are more general and could be applied to other combinations of dopants and hosts. \eu\ is a stable isotope (52\% natural abundance), and its collective enhancement of T-violation effects is estimated to be comparable to those of $^{225}$Ra and $^{223}$Fr \cite{Flambaum2020,Flambaum2020b}. Eu:YSO combines a number of advantages into one convenient platform: the intrinsic high sensitivity of $^{153}$Eu to nuclear T-violation, large numbers of strongly polarized atomic ions, and well-studied hyperfine structure that enables high-precision measurements on its nuclear spin states. In the following sections we describe the experimental measurement scheme and show how to use the properties of Eu:YSO to address challenges (i) and (ii). 

\section{Structure of Eu:YSO}\label{sec:structure}

YSO is a transparent crystal where the Y$^{3+}$ ions are located in non-centrosymmetric sites. Eu$^{3+}$ ions substitute for Y$^{3+}$ at two distinct $C_1$ sites: one with 7 adjacent oxygen ions (``site 1") and the other with 6 adjacent oxygen ions (``site 2'') \cite{Mirzai2021}. Each type of site, and therefore the electrical polarization of the corresponding Eu$^{3+}$ ion, can itself be oriented in a discrete set of directions within the crystal \cite{Ferrier2016}.

Eu:YSO has been extensively studied over several decades, primarily for its applications to quantum information storage and transduction \cite{Sellars2015, Timoney2012, Zhong2019}. In particular, narrow homogeneous and inhomogeneous linewidths have been observed on the $4f^6 \, {}^7F_0 \to 4f^6 \, {}^5D_0$ transition at 580 nm \cite{Konz2003}. The hyperfine and Zeeman interactions of the nuclear sublevels in the ${}^7F_0$ and ${}^5D_0$ states have been measured \cite{Cruzeiro2018}, optical pumping between the nuclear sublevels has been demonstrated \cite{Afzelius2012}, and long-lived coherence between these sublevels has been observed \cite{Arcangeli2015}. 

Linear Stark shifts of the $^7F_0$ $\rightarrow$ $^5D_0$ transition have been observed in Eu:YSO \cite{Zhang2020}. A linear Stark shift implies that Eu$^{3+}$ ions are statically polarized within the crystal. Static electric polarization of ions is not unusual -- in fact, this is what happens to the cations and anions in a polar molecule. But the difference between a gas-phase polar molecule and Eu:YSO is that gas-phase molecules exist in a quantum superposition of states with different orientations of their body-fixed dipole (i.e., rotational states), whereas Eu$^{3+}$ ions in Eu:YSO are projected into a classical statistical mixture of fixed orientations due to suppression of tunneling by the lattice. Eu:YSO therefore consists of a set of electrically polarized Eu$^{3+}$ ions, fixed in their orientation by the YSO crystal.

Despite being strongly electrically polarized and bonded to a set of neighboring $O^-$ ions as if in a polar molecule, Eu$^{3+}$ ions have an unusual property which is crucial to the measurement scheme described here: the lowest few electronic states in the ion retain their atom-like character. This property arises because the valence energy levels in Eu$^{3+}$, particularly the $^7F_0$ and $^5D_0$ states that are involved in the 580 nm optical transition, lie within the contracted and shielded $f$-shell \cite{Kaplyanskii2012,Zhang2020}. Furthermore, these states have zero electronic angular momenta, which drastically reduces their coupling to the lattice. The Eu:YSO system thus offers the combined benefits of atom-like narrow optical lines along with polar-molecule-like enhancement of T-violation effects.  

We parametrize the polarization of an Eu$^{3+}$ ion by its induced dipole moment $D \hat{n}$. We emphasize that this dipole moment is not the permanent electric dipole moment (EDM) that is measured in nuclear T-violation experiments. In \textit{non-centrosymmetric} sites, an electronic state of the ion can have a finite electric dipole moment $D$ due to odd-parity crystal field components and can thus exhibit linear Stark shifts \cite{Overhauser1958}. This effect leads to symmetric doublet splittings in optical transitions when two states with oppositely-oriented dipole moments (relative to the crystallographic axes) are degenerate in a non-centrosymmetric trapping site -- following the first observations in ruby crystals, this mechanism was named ``pseudo-Stark splitting" (PSS) \cite{Kaiser1961}. Similar to the description of Ref.\ \cite{Kaplyanskii2002}, in the case of the non-centrosymmetric $C_1$ trapping sites of YSO, the orientation of the dipole moment is fixed in an arbitrary direction relative to the crystallographic axes. However, within the overall \textit{centrosymmetric} YSO lattice ($C2/c$), the dipole moment is oriented relative to a pair of equivalent crystallographic directions, leading  to Eu$^{3+}$ ions with oppositely-oriented dipole moments in otherwise-physically-equivalent trapping sites in the macroscopic crystal.

The shift of the 580 nm transition in Eu:YSO in a lab electric field, $\Esca_\mrm{lab}$, is $\Delta E = - \Delta D \hat{n} \cdot \Evec_\mrm{lab}$, where $\Delta D = D({}^7F_0) - D({}^5D_0)$ is the differential dipole moment between the $^7F_0$ and $^5D_0$ electronic states. In the linear Stark shift measurements of Ref.\ \cite{Zhang2020}, an optically selected spectral class of ions prepared in $\Esca_\mrm{lab}=0$ split into two classes shifted by opposite amounts when $\Esca_\mrm{lab} \neq 0$ was applied. These measurements are consistent with the prior argument: that each unit cell in Eu:YSO has a static electric dipole moment $D \hat{n}$, and that the cells are found in two opposite orientations along each crystallographic axis in the macroscopic single crystal. The linear Stark shift coefficient $\Delta E = 27.1$ kHz/(V/cm) $\times \ \Esca_\mrm{lab}$ of the $^7F_0 \to {}^5D_0$ transition measured in Ref.\ \cite{Zhang2020}, for electric fields directed along the crystallographic $D1$ axis, indicates that the differential dipole moment along the $D1$ axis is $\Delta D = 0.02 \, e a_0$. As described below, the linear Stark shift of the optical transition provides a way to selectively address ions that have a specific direction of electric polarization $\hat{n}$. 

The effective Hamiltonian for the nuclear spin degrees of freedom, which accurately describes the hyperfine structure of $^{153}$Eu$^{3+}$ in YSO, is \begin{equation}
    \label{eq:hyperfine}
    \hat{H}_\mrm{eff} = \Bvec_\mrm{lab} \cdot \mathbf{M} \cdot \vec{I} + \vec{I} \cdot \mathbf{Q} \cdot \vec{I} - D \hat{n} \cdot \Evec_\mrm{lab} + W_T \vec{I} \cdot \hat{n}.
\end{equation}
Here $\vec{I}$ is the nuclear spin, $\Evec_\mrm{lab}$ is the lab electric field, and $\Bvec_\mrm{lab}$ is the lab magnetic field. $\mathbf{M}$ is the gyromagnetic tensor \cite{Smith2022}, such that $\vec{\mu} = \mathbf{M} \cdot \vec{I}$ is the magnetic moment. $\mathbf{Q}$ is the quadrupole tensor, and $D \hat{n}$ is the electric dipole moment of the Eu$^{3+}$ ion polarized by the crystal. The quantity $W_T$ parameterizes the nuclear T-violating energy shift. Since $\mathbf{M}$ and $\mathbf{Q}$ are influenced by crystal fields, they depend on the electronic state ($^7 F_0$ or $^5 D_0$), and they are anisotropic for the $C_1$ substitution sites where the $^{153}$Eu$^{3+}$ ions are trapped in Eu:YSO. The $\mathbf{M}$ tensor has been measured for $^{151}$Eu:YSO \cite{Longdell2006}, but not for the $^{153}$Eu isotope that is of interest here. Therefore, as a reasonable first approximation, we scaled the $\mathbf{M}$ tensor in Ref.\ \cite{Longdell2006} by the ratio of the nuclear magnetic moments of $^{153}$Eu to $^{151}$Eu \cite{Stone2005} for our calculations.

\begin{figure}
  \centering
  \includegraphics[width=\columnwidth]{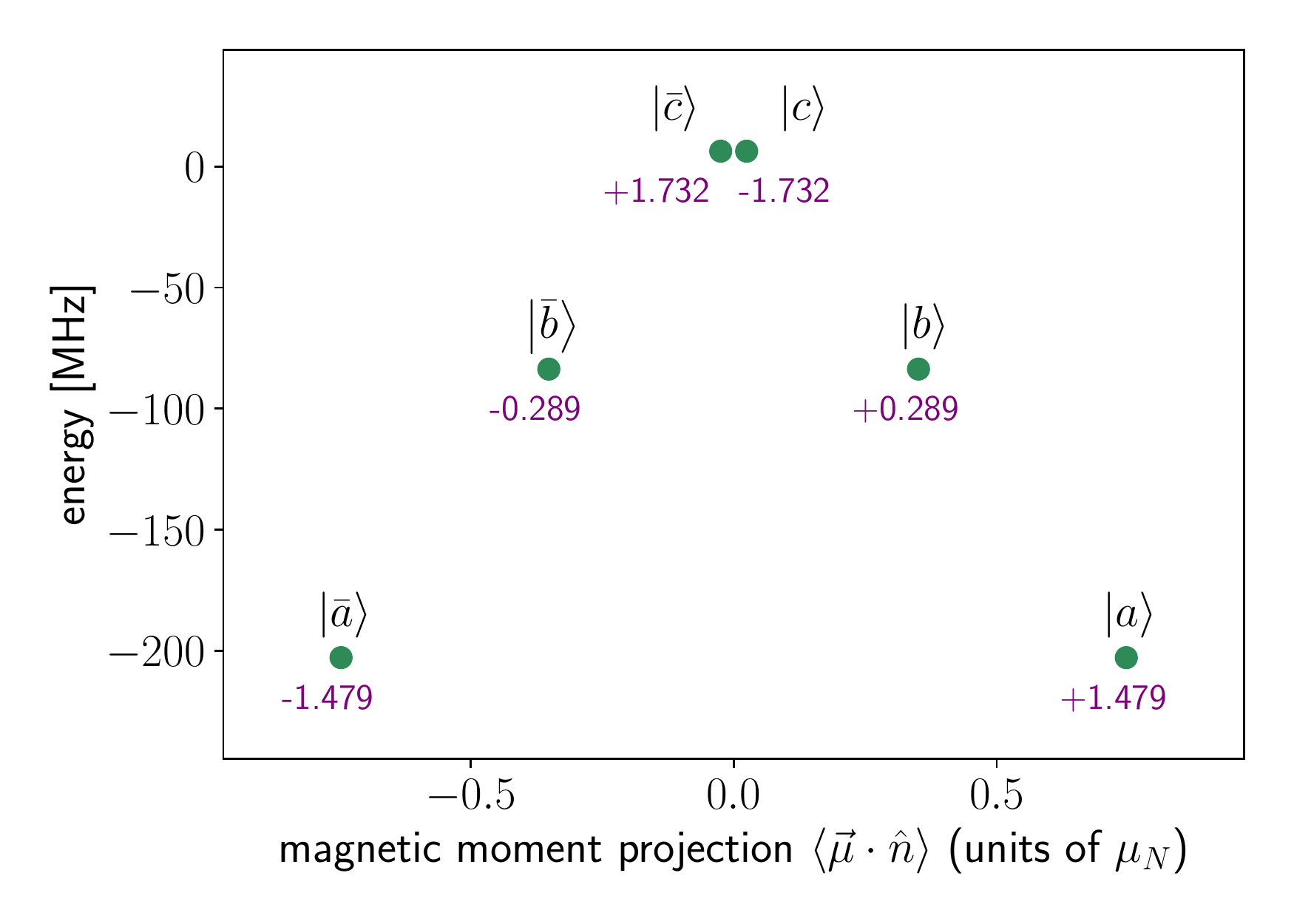}
  \caption{Energies, magnetic moment projections $\langle \vec{\mu} \cdot \hat{n} \rangle$, and nuclear T-violation sensitivity parameters $\zeta = \left\langle \vec{I} \cdot \hat{n} \right\rangle$ (purple text) for nuclear spin sublevels of the $^7F_0$ electronic state of $^{153}$Eu$^{3+}$ in YSO in zero magnetic field. These values were calculated using the constants measured in Refs.\ \cite{Yano1991, Longdell2006}, adjusted as described in the text. We fixed $\hat{n}$ parallel to the crystallographic $D1$ axis of YSO for this calculation.}
  \label{fig:states}
\end{figure}

The eigenstates of this Hamiltonian for the $^7F_0$ electronic state in the absence of external fields (and with $W_T = 0$) are represented as points in the diagram in Fig.\ \ref{fig:states}. 
The eigenstates appear in degenerate pairs that are related to each other by a T-transformation, as expected from Kramers' degeneracy theorem \cite{Klein1952}. We denote these Kramers pairs as $\bar{a},a$, $\bar{b},b$ and $\bar{c},c$ in increasing order of energy. We use this nomenclature for the eigenstates instead of $m_I$ labels that appear in the literature, since $m_I$ is not a good (or particularly useful) quantum number when there is no azimuthal symmetry. Also shown in Fig. \ref{fig:states} are the sensitivity coefficients $\zeta = \avg{\vec{I} \cdot \hat{n}}$ in each state --- these values are proportional to the energy shift of the state if T-violation were present. T-violating energy shifts have equal magnitude and opposite sign for the two states in a Kramers pair. So in a nutshell, the experiment involves searching for the breakdown of Kramers' degeneracy due to T-violation. 

Note, however, that the states in a Kramers pair have opposite signs of magnetic moments as well. Therefore, the experiment needs to be able to distinguish T-violating energy shifts from mundane Zeeman shifts due to magnetic fields. This is easily accomplished: the $\zeta$ values for any given state have equal magnitude, but \textit{opposite signs}, in ions with opposite values of $\hat{n}$ (i.e., ions with opposite electric polarization). So the experiment consists of measuring energy differences between a Kramers pair in ions that are polarized in a particular direction in the lattice, and comparing the result with the same measurement in ions polarized in the opposite direction. For definiteness, the measurement scheme in Section \ref{sec:scheme} focuses on measuring the $T$-violating energy difference between the states labelled as $\ket{\bar{a}}$ and $\ket{a}$ in Fig.\ \ref{fig:states}.

A key step in the measurement scheme is the transfer of population into specific Kramers pairs through optical pumping. Note that, at any given optical frequency within the inhomogeneous optical line, ions from many different spectral classes (i.e., local crystal environments) can be resonant with the laser. Each spectral class is associated with nine distinct optical frequencies, from the combinations of three Kramers pairs each in $^7F_0$ and $^5D_0$. 

In contrast with gas-phase atoms, Eu:YSO can be optically pumped into selected hyperfine states despite the lack of electronic angular momentum in $^7F_0$ and $^5D_0$. The main reason is that the quadrupole tensor $\mathbf{Q}$ is different in the $^7F_0$ and $^5D_0$ states, and so the hyperfine eigenstates in these two electronic states have different orientations of $\vec{I}$ relative to the crystallographic axes. Therefore, the absorption or spontaneous emission of a photon on the $^7F_0 - {}^5D_0$ transition can reorient $\vec{I}$, resulting in finite optical transition matrix elements between any $^7F_0$ hyperfine state $\ket{i}$ and $^5D_0$ hyperfine state $\ket{j_e}$. These optical transition matrix elements are listed in Table \ref{tab:transition_rates}. Applying optical pulses with appropriate sidebands to address multiple states, it is possible to selectively populate the $\bar{b},b$ Kramers pair in $^7F_0$.

\begin{table}[]
    \centering
    \begin{tabular}{lrrrrrr}
        \toprule
        {} &      $\ket{a}$ &     $\ket{\bar{a}}$ &      $\ket{b}$ &     $\ket{\bar{b}}$ &      $\ket{c}$ &     $\ket{\bar{c}}$ \\
        \midrule
        $\bra{a_e}$  &  0.031 &  0.001 &  0.157 &  0.012 &  0.799 &  0.000 \\
        $\bra{\bar{a}_e}$ &  0.001 &  0.031 &  0.012 &  0.157 &  0.000 &  0.799 \\
        $\bra{b_e}$  &  0.094 &  0.004 &  0.690 &  0.026 &  0.163 &  0.022 \\
        $\bra{\bar{b}_e}$ &  0.004 &  0.094 &  0.026 &  0.690 &  0.022 &  0.163 \\
        $\bra{c_e}$  &  0.866 &  0.004 &  0.105 &  0.009 &  0.007 &  0.009 \\
        $\bra{\bar{c}_e}$ &  0.004 &  0.866 &  0.009 &  0.105 &  0.009 &  0.007 \\
        \bottomrule
    \end{tabular}
    \caption{Relative transition probabilities $|\braket{j_e}{i}|^2$ between hyperfine sublevels of $^7F_0 \rightarrow {}^5D_0$ in \eu:YSO in zero magnetic field, calculated using the constants measured in Refs. \cite{Yano1991, Yano1992}. Hyperfine levels belonging to the excited electronic state are denoted with the subscript ``$e$''.}
    \label{tab:transition_rates}
\end{table}

\section{Measurement scheme}\label{sec:scheme}
\begin{figure*}
    \centering
    \includegraphics[width=\textwidth]{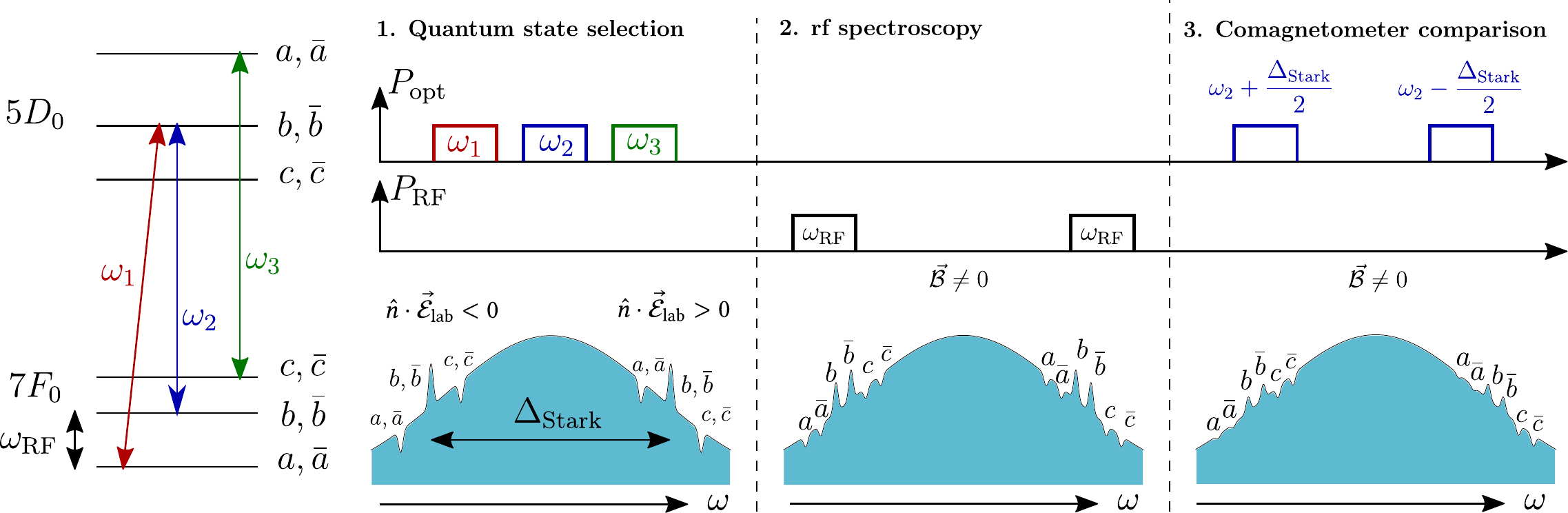}
    \caption{Overview of the measurement scheme. The $^7F_0 - {}^5D_0$ optical transition in Eu:YSO is used to optically pump the nuclear spin sublevels. In Step 1, laser sidebands tuned to different hyperfine transitions are used to deplete the $a,\bar{a}$ and $c,\bar{c}$ states -- creating holes in the inhomogeneously broadened optical line -- and to populate the $b,\bar{b}$ states, which creates antiholes \cite{Macfarlane1987}. Applying a lab electric field moves the holes and antiholes to different spectral neighborhoods as shown in the diagram, depending on the relative orientation between $\Evec_\mrm{lab}$ and the dopant ion's electrical polarization $\hat{n}$. In Step 2, rf pulses are used to drive the $\bar{b}-\bar{a}$ and $b-a$ transitions. In Step 3 the populations excited into $a,\bar{a}$ are detected using optical absorption, and the rf resonance frequencies are compared for opposite orientations of $\hat{n}$. A similar method can be used for measuring the $\bar{b}-\bar{c}$ and $b-c$ transition.}
    \label{fig:expt_steps}
\end{figure*}

The measurement sequence, illustrated in Fig.\ \ref{fig:expt_steps}, is as follows.
\begin{enumerate}[1.]
    \item \textbf{State selection.} The 580 nm $^7F_0 - {}^5D_0$ optical transition of Eu$^{3+}$, with appropriately chosen rf sidebands, is used to deplete the $\ket{a},\ket{\bar{a}}$ and $\ket{c},\ket{\bar{c}}$ states, and pump the population into $\ket{b},\ket{\bar{b}}$. 
    
    \item \textbf{rf spectroscopy.} In order to precisely measure the T-violating energy shift, the $\bar{b}-\bar{a}$ and $b-a$ transitions are driven using a sequence of rf pulses. 

    \item \textbf{Comagnetometer comparison.} Populations transferred to $a,\bar{a}$ by the rf pulses are detected using optical absorption, in order to measure the rf resonance frequencies. The resonances measured in this way are compared for ions with opposite values of $\hat{n}$, taking advantage of the Stark shifts of the optical transitions. 
\end{enumerate}

In Step 1, Eu$^{3+}$ ions are initialized in the $\bar{b},b$ states using a sequence of laser pulses. In gas-phase atoms, two lasers (one to deplete $\bar{a},a$ and one to deplete $\bar{c},c$) would be sufficient to transfer population to the $\bar{b},b$ Kramers pair. But due to inhomogeneous broadening of the ${}^7F_0 - {}^5D_0$ optical transition in Eu:YSO, these two lasers also \emph{depopulate} the desired $\bar{b},b$ Kramers pair in spectral classes whose optical resonances are shifted by linear combinations of the ${}^7F_0$ and ${}^5D_0$ hyperfine intervals. Therefore three different optical frequencies are necessary to ensure that one spectral hole (or anti-hole) is uniquely associated with one Kramers pair. The illustration for Step 1 in Fig. \ref{fig:expt_steps} shows these laser frequencies. A detailed description of the hyperfine state-preparation protocol can be found in Ref.\ \cite{Afzelius2012}.

In Step 2, a small bias magnetic field $\Bsca_\mrm{lab}$ is applied for experimental convenience, so that nonzero frequencies can be measured. rf spectroscopy of the $\bar{a}-\bar{b}$ and $a-b$ transitions is then used to measure the frequency difference between the $\bar{a},a$ states, $\nu_{\bar{a}a}$. A nonzero value of $\nu_{\bar{a}a}$ arises due to the Zeeman shift from the bias field, plus a putative shift due to the presence of T-violation. 

The main obstacle to precision rf spectroscopy of $\nu_{\bar{a}a}$ is the inhomogeneous broadening of the $\bar{a}-\bar{b}$ and $a-b$ rf transitions. On the face of it, the precision of $\nu_{\bar{a}a}$ would seem to scale as $1 / \sqrt{T_2}$ with the ensemble coherence time $T_2 \sim 1/{\Gamma_\mathrm{inh}}$ arising from the inhomogeneous width $\Gamma_\mathrm{inh}$ of the $\bar{a}-\bar{b}$ and $a-b$ transitions. Although Eu:YSO would nonetheless offer competitive T-violation constraints with $T_2 \sim 2.3 \, \mu$s \cite{Timoney2012}, it is in fact possible to obtain much better precision, scaling inversely with the \textit{hyperfine coherence time} $\tau = 15$ ms \cite{Macfarlane2014,Arcangeli2014,Arcangeli2015}.

The key to measuring the small frequency difference $\nu_{\bar{a}a}$, despite its being masked by inhomogeneous broadening, is to use coherent quantum beat (CQB) spectroscopy \cite{Ramachandran2023}. The idea behind the CQB method is summarized in the Appendix \ref{sec:cqb}, and further details and numerical results can be found in Ref.\ \cite{Ramachandran2023}. In Step 2, applying CQB to the four-level system -- $b, \bar{b}$ and $a, \bar{a}$ -- allows for simultaneous measurements of $\nu_{\bar{a}a}$ on the entire ensemble of Eu$^{3+}$ nuclear spins prepared in Step 1, leading to larger signals and longer coherence time. Inhomogeneous broadening of the hyperfine transitions is not a fundamental barrier to precision measurements of T-violation in solid-state systems. 

In Step 3, we take advantage of the $^7F_0 - {}^5D_0$ transition to optically read out the populations in the ${\bar{b}}, b, {\bar{a}}, a$ states, using absorption or fluorescence spectroscopy. Crucially, these populations can be \emph{separately} measured for Eu$^{3+}$ ions polarized in opposite directions. Due to the linear Stark shift and the sub-kHz homogeneous linewidth of the $^7F_0 - {}^5D_0$ transition, a small lab electric field, $\Esca_\mrm{lab} \approx$ 5 V/cm, suffices to widely separate the optical transitions in ions with opposite values of $\hat{n}$. These oppositely polarized ions then serve as ideal ``co-magnetometers'', with properties that are completely identical in every way except for the sign of the T-violating energy shift.

\subsection{Suppression of systematic errors}
During the measurement, spectroscopy of the $\nu_{\bar{a}a}$ can be repeated while varying a number of experimental parameters. These parameters include, but are not limited to, the static magnetic and electric field strengths, rf pulse amplitudes and phases, the volume illuminated by the laser within the crystal, and the spectral class of Eu$^{3+}$ ions that are optically state-selected. These variations can be augmented with a number of ``reversals'', i.e. discrete switches, of parameters such as the signs of the static electric and magnetic fields. The reversals modulate the energy shifts from a true P-odd T-odd nuclear moment (proportional to $\zeta$) in a characteristic way, allowing us to separate T-violation signals from spurious systematics.

A useful comparison for distinguishing true T-violation in such measurements uses the ``co-magnetometer'' ions mentioned above. This process is analogous to the effect used to great advantage in electron EDM experiments with polar molecules \cite{DeMille2000}. Additionally, since the ion is already strongly polarized within the crystal, variations in the small applied electric field $\Esca_\mrm{lab}$ do not appreciably change the T-violating energy shift. Thus the value of $\zeta$ is essentially unaffected by variations in the magnitude of $\Esca_\mrm{lab}$. On the other hand many classes of systematic errors, e.g. due to non-reversing electric fields or leakage currents caused by the fields, strongly depend on $\Esca_\mrm{lab}$. Therefore experiments conducted with different values of $\Esca_\mrm{lab}$ allow a large variety of systematic errors to be distinguished from genuine signals.

Yet another useful switch is to interleave measurements on the $\bar{b}-\bar{a}, b-a$ transitions with measurements on other pairs such as $\bar{b}-\bar{c}, b-c$ (see Fig.\ \ref{fig:states}) under otherwise identical conditions. These comparisons, between sets of states with very different values of $\zeta$ and different magnetic moments, will affect a true T-violation signal in a characteristic and calculable way, distinct from errors due to uncompensated magnetic fields.

Finally, another convenient switch is offered by comparisons between measurements on $^{151}$Eu$^{3+}$ and $^{153}$Eu$^{3+}$ isotopes under otherwise identical measurement parameters. These ions have the same nuclear spin ($I=5/2$), nearly equal natural isotopic abundance and comparable atomic and magnetic moments, but only $^{153}$Eu has the nuclear octupole deformation which enhances the effects of microscopic T-violation \cite{Flambaum2020}. Therefore the two isotopes offer one more useful co-magnetometer-like comparison. Using the shift in the hyperfine Hamiltonian between these isotopes, a simple change in the laser frequencies used in Step 1 above allows switching between the isotopes within the same crystal, in order to separate T-violation signals from systematics. 

Table \ref{tab:systematics} summarizes some of the systematic error controls available with this combination of experimental platform and measurement scheme.

\begin{table*}[]
    \centering
    \begin{tabular}{lll}
     Systematic effect & Detected and suppressed using \\
     \hline 
     B-field drifts & $\hat{n}$ switch, spin states switch, isotope switch \\
     Non-reversing B-fields & $\Esca_\mrm{lab}$ magnitude, $\hat{n}$ switch, spin states switch, isotope switch \\
     Non-reversing E-fields & $\Esca_\mrm{lab}$ magnitude, spin states switch, isotope switch \\
     Laser light shifts & $\Esca_\mrm{lab}$ magnitude, $\hat{n}$ switch, spin states switch, isotope switch \\
     rf light shifts/pulse errors & $\Bvec_\mrm{lab}$ reversal, $\hat{n}$ switch, isotope switch, rf pulse magnitude and width \\
     Phonon effects & above switches + crystal temperature variation \\
     Crystal impurities/defects  &  differently doped/enriched crystals\\
     \hline
    \end{tabular}
    \caption{Potential systematics and paths to their suppression in the experiment. The technical simplicity of the experimental system, coupled with the many useful controls available in Eu:YSO, enables T-violating energy shifts to be cleanly distinguished from mimicking systematics.}
    \label{tab:systematics}
\end{table*}

The Eu:YSO system has a number of practical advantages for experiments. The crystal occupies a small volume, making it conveniently easy to shield from background electromagnetic fields. It is simple to operate experiments with multiple crystals (e.g., with different doping concentration, or different isotopic enhancements) within the measurement volume or to change the temperature of the crystal. A single laser, with appropriate rf sidebands imprinted by acousto- or electro-optic modulators, is adequate for the state-preparation and readout steps. The crucial spectroscopy step is done entirely using rf spectroscopy, which allows for precise control of the pulse amplitudes and phases. 

\section{Estimated sensitivity to new physics}
The essence of the experiment is to search for a T-violating energy shift between different orientations of the $^{153}$Eu nucleus within a polarized Eu$^{3+}$ ion. The sensitivity of this process to T-violation is determined by (i) the polarization of the Eu$^{3+}$ ion by the crystal, which results in T-violating energy shifts, (ii) the intrinsic sensitivity of the $^{153}$Eu nucleus to fundamental T-violation parameters, and (iii) the statistical precision with which the T-violating shift is measured.

In the following, we will estimate the polarization of the Eu$^{3+}$ ions in Eu:YSO using published experimental measurements and theory calculations. First, we introduce a quantity that we denote as the crystal electric field, $\vec{\mathcal{E}}_\mathrm{xtl}$. This is the field that one would have to apply to a free-space ion to obtain the induced dipole moment $D\hat{n}$ of the Eu$^{3+}$ ions in Eu:YSO. The magnitude of $\mathcal{E}_\mathrm{xtl}$ can be estimated from the measured linear Stark shift of the $^7F_0 - {}^5D_0$ transition. 

The linear shift measured in experiments arises from an underlying quadratic shift that involves the crystal electric field and the atomic polarizabilities: $\Delta E = \Delta \alpha \left(\vec{\mathcal{E}}_{\mathrm{xtl}} + \Evec'_{\mathrm{lab}} \right)^2$, where $\Esca'_\mathrm{lab} = \Esca_\mrm{lab}/\epsilon_r$ is the lab field shielded by the permittivity $\epsilon_{r} \approx 10$ \cite{Carvalho2015} of the crystal. The atomic polarizability difference between the  $^5D_0$ and $^7F_0$ states is $\Delta \alpha = \alpha_e - \alpha_g$. Therefore $\Delta \nu_E = 2 \Delta \alpha \, \Evec_\mrm{xtl} \cdot \Evec'_\mrm{lab}$. 

In order to estimate $\Esca_\mrm{xtl}$ from the measured Stark shift coefficient $\Delta \nu_E \sim 27$ kHz/(V/cm) \cite{Zhang2020}, we need estimates of $\Delta \alpha$. The differential polarizability of the states in Eu:YSO is presently unknown, although the polarizability of the ground state of the ion is known to be $\alpha_g = 1.1$ a.u. in free space \cite{Clavaguera2005}. In order to conservatively account for the higher polarizability of the ion within the YSO lattice \cite{Nishimura1991}, and for the unknown (although likely much smaller \cite{Nishimura1991}) polarizability of the $^5D_0$ state, we assume in the worst-case that $\Delta \alpha = 100$ atomic units. The resulting estimate of the crystal electric field is  $\mathcal{E}_\mathrm{xtl} \gtrsim 1.3 \ \mathrm{MV/cm}$. 

Next, the T-violating energy shift measured in experiments can be related to a microscopic measure of T-violation. We write the T-violating shift discussed in Section \ref{sec:scheme} as $\nu_{\bar{a}a} = 2 \zeta_a \,  \kappa \,  \theta \, \Esca_\mrm{xtl}$, where $\zeta_a$ is the absolute value of the $\zeta$ parameter in the $\bar{a}, a$ states. The quantity $\theta$ is the equivalent value of the quantum chromodynamics $\bar{\theta}_\mrm{QCD}$ parameter that would lead to the same physical effects, and is thus a convenient dimensionless way to quantify T-violation due to new physics. The value of $\kappa$ has been calculated for $^{153}$Eu$^{3+}$ in Ref.\ \cite{Flambaum2020b} to be $\kappa({}^{153}\mrm{Eu}) = -1.2 \times 10^{-17}$ $e$ cm. Thus we arrive at an estimate of the T-violating shift in Eu:YSO, which is $\nu_{\bar{a}a} \sim 1.1 \times 10^4 \, \bar{\theta}$ Hz.  

The above number is based on a highly conservative estimate of $\Esca_\mrm{xtl}$. More accurate estimates of the new physics and dark matter sensitivities of the Eu:YSO system will need to be supported by accurate calculations of the electrical polarization of Eu$^{3+}$ in non-centrosymmetric sites \footnote{A calculation for a similar system, EuCl$_3$, has been presented in Ref. \cite{Sushkov2023}.}.

The experimental precision in measuring the T-violating shift depends on three factors: the number of Eu$^{3+}$ ions involved in a measurement, $N$, the nuclear spin coherence time, $\tau$, and the total integration time, $T_\mrm{int}$.  Assuming an Eu concentration of 0.01\%, and a 1-mm-diameter laser beam propagating through a 10-mm-long crystal, we estimate the number of ions resonant with the laser to be $N=10^{14}$ (accounting for the fraction of ions addressed in the optical state-selection process). As mentioned above and discussed in Ref.\ \cite{Ramachandran2023}, coherent quantum beat spectroscopy allows measurements of $\nu_{\bar{a}a}$ with a coherence time that is not limited by the ensemble $T_2$ for nuclear spin transitions. For the estimate of new physics sensitivity, we use $\tau=15$ ms  measured at 4 K in Refs.\ \cite{Macfarlane2014,Arcangeli2014,Arcangeli2015}. \\

The precision in the measurement of the T-violating shift is
\begin{equation}\label{eq:sensitivity}
    \delta \nu = \frac{1}{2 \mathscr{S}} \, \sqrt{\frac{1+ \beta}{\tau T_\mrm{int}}}
\end{equation}
where $\mathscr{S}$ is the signal-to-noise ratio of population measurements that can be obtained within a measurement time $T_m = \beta \tau$. Here $1+\beta$ is a duty-cycle factor that encapsulates how efficiently the measurements use the available coherence time. We assume that the full extent of the signal-to-noise from $N=10^{14}$ ions, $\mathscr{S} = 10^7$, can be obtained in a measurement time $T_m = 2\tau$ (i.e., $\beta=2$), using low-noise laser absorption measurements of the hyperfine state populations. These estimates suggest a statistical sensitivity $\delta \nu \sim 2.4 \ \mrm{nHz} \times \sqrt{\frac{1 \ \mrm{day}}{T_\mrm{int}}}$. 

Combining the estimate of T-violating shift in the Eu:YSO system, $\nu_{\bar{a}a}$, with the precision obtainable from the measurement scheme, $\delta \nu$, we find that the resulting T-violation sensitivity is $\delta \theta \lesssim  3 \times 10^{-13} \sqrt{\frac{1 \ \mrm{day}}{T_\mrm{int}}}$. The realization of the scheme developed in this paper could lead to a considerable improvement over current bounds on T-violation \cite{Abel2020,Graner2016}. 

We note that nuclear spin coherence for thousands of seconds has been demonstrated in the Eu:YSO system at high magnetic fields \cite{Longdell2006,Sellars2015}. While we have conservatively made our estimate using the much smaller value of $\tau$ in low magnetic fields reported for Eu:YSO in the literature, it is enticing to anticipate new measurement techniques that could potentially unlock much longer coherence times for T-violation measurements in Eu:YSO.

In addition to P-odd, T-odd nuclear moments, T-violation can also appear due to time-dependent phenomena, such as oscillating electric dipole moments arising from wave-like dark matter \cite{Graham2013,Roussy2021,Aybas2021}. Any such oscillations need to be measured with a bandwidth higher than two inverse periods, else they average away to zero. The high sensitivity obtainable within a short integration time using Eu:YSO leads to useful bandwidth: we expect to be sensitive to fluctuations $\delta \theta < 10^{-10}$ within one second of integration, enabling a broadband search for wave-like dark matter over the mass range from $10^{-14} - 10^{-18}$ eV. \\

\section{Conclusion}
We have described a means to measure T-violation with improved sensitivity, using octupole-deformed nuclei doped into non-centrosymmetric sites in a crystal. Our approach leverages the high intrinsic T-violation sensitivity of octupole-deformed nuclei, while using electrically polarized ions that are deeply trapped within a crystal. The measurement can be performed using a simple and compact experimental system. In order to match the enhanced precision available in this system with the required level of systematic error control, we have identified a number of tests and reversals to isolate genuine new physics from spurious backgrounds. Although a low symmetry solid-state system may seem to be a complex place for precision measurements, we have shown how the unique properties of Eu:YSO paired with an appropriate measurement scheme can yield high sensitivity to T-violation. 

~\\
\textit{Acknowledgments. --} We thank David DeMille, Andrew Jayich, Jonathan Weinstein, Yoshiro Takahashi, Mingyu Fan and Julia Ford for helpful conversations and advice. HDR acknowledges support from an NSERC Canada Graduate Scholarship. ACV acknowledges support from the Canada Research Chairs program. This work was supported by an NSERC Discovery Grant.

\bibliography{eu}

\providecommand{\noopsort}[1]{}\providecommand{\singleletter}[1]{#1}%
\begin{thebibliography}{49}%
\makeatletter
\providecommand \@ifxundefined [1]{%
 \@ifx{#1\undefined}
}%
\providecommand \@ifnum [1]{%
 \ifnum #1\expandafter \@firstoftwo
 \else \expandafter \@secondoftwo
 \fi
}%
\providecommand \@ifx [1]{%
 \ifx #1\expandafter \@firstoftwo
 \else \expandafter \@secondoftwo
 \fi
}%
\providecommand \natexlab [1]{#1}%
\providecommand \enquote  [1]{``#1''}%
\providecommand \bibnamefont  [1]{#1}%
\providecommand \bibfnamefont [1]{#1}%
\providecommand \citenamefont [1]{#1}%
\providecommand \href@noop [0]{\@secondoftwo}%
\providecommand \href [0]{\begingroup \@sanitize@url \@href}%
\providecommand \@href[1]{\@@startlink{#1}\@@href}%
\providecommand \@@href[1]{\endgroup#1\@@endlink}%
\providecommand \@sanitize@url [0]{\catcode `\\12\catcode `\$12\catcode
  `\&12\catcode `\#12\catcode `\^12\catcode `\_12\catcode `\%12\relax}%
\providecommand \@@startlink[1]{}%
\providecommand \@@endlink[0]{}%
\providecommand \url  [0]{\begingroup\@sanitize@url \@url }%
\providecommand \@url [1]{\endgroup\@href {#1}{\urlprefix }}%
\providecommand \urlprefix  [0]{URL }%
\providecommand \Eprint [0]{\href }%
\providecommand \doibase [0]{https://doi.org/}%
\providecommand \selectlanguage [0]{\@gobble}%
\providecommand \bibinfo  [0]{\@secondoftwo}%
\providecommand \bibfield  [0]{\@secondoftwo}%
\providecommand \translation [1]{[#1]}%
\providecommand \BibitemOpen [0]{}%
\providecommand \bibitemStop [0]{}%
\providecommand \bibitemNoStop [0]{.\EOS\space}%
\providecommand \EOS [0]{\spacefactor3000\relax}%
\providecommand \BibitemShut  [1]{\csname bibitem#1\endcsname}%
\let\auto@bib@innerbib\@empty
\bibitem [{\citenamefont {Sakharov}(1991)}]{Sakharov1991}%
  \BibitemOpen
  \bibfield  {author} {\bibinfo {author} {\bibfnamefont {A.~D.}\ \bibnamefont
  {Sakharov}},\ }\bibfield  {title} {\bibinfo {title} {Violation of {$CP$}
  invariance, {$C$} asymmetry, and baryon asymmetry of the universe},\ }\href
  {https://doi.org/10.1070/pu1991v034n05abeh002497} {\bibfield  {journal}
  {\bibinfo  {journal} {Sov. Phys. Uspekhi}\ }\textbf {\bibinfo {volume}
  {34}},\ \bibinfo {pages} {392} (\bibinfo {year} {1991})}\BibitemShut
  {NoStop}%
\bibitem [{\citenamefont {Engel}\ \emph {et~al.}(2013)\citenamefont {Engel},
  \citenamefont {Ramsey-Musolf},\ and\ \citenamefont {{van
  Kolck}}}]{Engel2013}%
  \BibitemOpen
  \bibfield  {author} {\bibinfo {author} {\bibfnamefont {J.}~\bibnamefont
  {Engel}}, \bibinfo {author} {\bibfnamefont {M.~J.}\ \bibnamefont
  {Ramsey-Musolf}},\ and\ \bibinfo {author} {\bibfnamefont {U.}~\bibnamefont
  {{van Kolck}}},\ }\bibfield  {title} {\bibinfo {title} {Electric dipole
  moments of nucleons, nuclei, and atoms: The standard model and beyond},\
  }\href@noop {} {\bibfield  {journal} {\bibinfo  {journal} {Progress in
  Particle and Nuclear Physics}\ }\textbf {\bibinfo {volume} {71}},\ \bibinfo
  {pages} {21} (\bibinfo {year} {2013})}\BibitemShut {NoStop}%
\bibitem [{\citenamefont {DeMille}\ \emph {et~al.}(2017)\citenamefont
  {DeMille}, \citenamefont {Doyle},\ and\ \citenamefont
  {Sushkov}}]{DeMille2017}%
  \BibitemOpen
  \bibfield  {author} {\bibinfo {author} {\bibfnamefont {D.}~\bibnamefont
  {DeMille}}, \bibinfo {author} {\bibfnamefont {J.~M.}\ \bibnamefont {Doyle}},\
  and\ \bibinfo {author} {\bibfnamefont {A.~O.}\ \bibnamefont {Sushkov}},\
  }\bibfield  {title} {\bibinfo {title} {Probing the frontiers of particle
  physics with tabletop-scale experiments},\ }\href@noop {} {\bibfield
  {journal} {\bibinfo  {journal} {Science}\ }\textbf {\bibinfo {volume}
  {357}},\ \bibinfo {pages} {990} (\bibinfo {year} {2017})}\BibitemShut
  {NoStop}%
\bibitem [{\citenamefont {Sushkov}\ \emph {et~al.}(1984)\citenamefont
  {Sushkov}, \citenamefont {Flambaum},\ and\ \citenamefont
  {Khriplovich}}]{Sushkov1984}%
  \BibitemOpen
  \bibfield  {author} {\bibinfo {author} {\bibfnamefont {O.~P.}\ \bibnamefont
  {Sushkov}}, \bibinfo {author} {\bibfnamefont {V.~V.}\ \bibnamefont
  {Flambaum}},\ and\ \bibinfo {author} {\bibfnamefont {I.~B.}\ \bibnamefont
  {Khriplovich}},\ }\bibfield  {title} {\bibinfo {title} {Possibility of
  investigating {P}- and {T}-odd nuclear forces in atomic and molecular
  experiments},\ }\href@noop {} {\bibfield  {journal} {\bibinfo  {journal}
  {Sov. Phys. - JETP}\ }\textbf {\bibinfo {volume} {60}} (\bibinfo {year}
  {1984})}\BibitemShut {NoStop}%
\bibitem [{\citenamefont {Skripnikov}\ \emph {et~al.}(2020)\citenamefont
  {Skripnikov}, \citenamefont {Mosyagin}, \citenamefont {Titov},\ and\
  \citenamefont {Flambaum}}]{Skripnikov2020}%
  \BibitemOpen
  \bibfield  {author} {\bibinfo {author} {\bibfnamefont {L.~V.}\ \bibnamefont
  {Skripnikov}}, \bibinfo {author} {\bibfnamefont {N.~S.}\ \bibnamefont
  {Mosyagin}}, \bibinfo {author} {\bibfnamefont {A.~V.}\ \bibnamefont
  {Titov}},\ and\ \bibinfo {author} {\bibfnamefont {V.~V.}\ \bibnamefont
  {Flambaum}},\ }\bibfield  {title} {\bibinfo {title} {Actinide and lanthanide
  molecules to search for strong {CP}-violation},\ }\href@noop {} {\bibfield
  {journal} {\bibinfo  {journal} {Phys. Chem. Chem. Phys.}\ }\textbf {\bibinfo
  {volume} {22}},\ \bibinfo {pages} {18374} (\bibinfo {year}
  {2020})}\BibitemShut {NoStop}%
\bibitem [{\citenamefont {Sandars}(1967)}]{Sandars1967}%
  \BibitemOpen
  \bibfield  {author} {\bibinfo {author} {\bibfnamefont {P.~G.~H.}\
  \bibnamefont {Sandars}},\ }\bibfield  {title} {\bibinfo {title}
  {Measurability of the proton electric dipole moment},\ }\href
  {https://doi.org/10.1103/PhysRevLett.19.1396} {\bibfield  {journal} {\bibinfo
   {journal} {Phys. Rev. Lett.}\ }\textbf {\bibinfo {volume} {19}},\ \bibinfo
  {pages} {1396} (\bibinfo {year} {1967})}\BibitemShut {NoStop}%
\bibitem [{\citenamefont {Wilkening}\ \emph {et~al.}(1984)\citenamefont
  {Wilkening}, \citenamefont {Ramsey},\ and\ \citenamefont
  {Larson}}]{Wilkening1984}%
  \BibitemOpen
  \bibfield  {author} {\bibinfo {author} {\bibfnamefont {D.~A.}\ \bibnamefont
  {Wilkening}}, \bibinfo {author} {\bibfnamefont {N.~F.}\ \bibnamefont
  {Ramsey}},\ and\ \bibinfo {author} {\bibfnamefont {D.~J.}\ \bibnamefont
  {Larson}},\ }\bibfield  {title} {\bibinfo {title} {Search for {$P$} and {$T$}
  violations in the hyperfine structure of thallium fluoride},\ }\href
  {https://doi.org/10.1103/PhysRevA.29.425} {\bibfield  {journal} {\bibinfo
  {journal} {Phys. Rev. A}\ }\textbf {\bibinfo {volume} {29}},\ \bibinfo
  {pages} {425} (\bibinfo {year} {1984})}\BibitemShut {NoStop}%
\bibitem [{\citenamefont {Isaev}\ \emph {et~al.}(2010)\citenamefont {Isaev},
  \citenamefont {Hoekstra},\ and\ \citenamefont {Berger}}]{Isaev2010}%
  \BibitemOpen
  \bibfield  {author} {\bibinfo {author} {\bibfnamefont {T.~A.}\ \bibnamefont
  {Isaev}}, \bibinfo {author} {\bibfnamefont {S.}~\bibnamefont {Hoekstra}},\
  and\ \bibinfo {author} {\bibfnamefont {R.}~\bibnamefont {Berger}},\
  }\bibfield  {title} {\bibinfo {title} {Laser-cooled raf as a promising
  candidate to measure molecular parity violation},\ }\href
  {https://doi.org/10.1103/PhysRevA.82.052521} {\bibfield  {journal} {\bibinfo
  {journal} {Phys. Rev. A}\ }\textbf {\bibinfo {volume} {82}},\ \bibinfo
  {pages} {052521} (\bibinfo {year} {2010})}\BibitemShut {NoStop}%
\bibitem [{\citenamefont {Grasdijk}\ \emph {et~al.}(2021)\citenamefont
  {Grasdijk}, \citenamefont {Timgren}, \citenamefont {Kastelic}, \citenamefont
  {Wright}, \citenamefont {Lamoreaux}, \citenamefont {DeMille}, \citenamefont
  {Wenz}, \citenamefont {Aitken}, \citenamefont {Zelevinsky}, \citenamefont
  {Winick} \emph {et~al.}}]{Grasdijk2021}%
  \BibitemOpen
  \bibfield  {author} {\bibinfo {author} {\bibfnamefont {J.~O.}\ \bibnamefont
  {Grasdijk}}, \bibinfo {author} {\bibfnamefont {O.}~\bibnamefont {Timgren}},
  \bibinfo {author} {\bibfnamefont {J.}~\bibnamefont {Kastelic}}, \bibinfo
  {author} {\bibfnamefont {T.}~\bibnamefont {Wright}}, \bibinfo {author}
  {\bibfnamefont {S.~K.}\ \bibnamefont {Lamoreaux}}, \bibinfo {author}
  {\bibfnamefont {D.~P.}\ \bibnamefont {DeMille}}, \bibinfo {author}
  {\bibfnamefont {K.}~\bibnamefont {Wenz}}, \bibinfo {author} {\bibfnamefont
  {M.}~\bibnamefont {Aitken}}, \bibinfo {author} {\bibfnamefont
  {T.}~\bibnamefont {Zelevinsky}}, \bibinfo {author} {\bibfnamefont
  {T.}~\bibnamefont {Winick}}, \emph {et~al.},\ }\bibfield  {title} {\bibinfo
  {title} {{CeNTREX}: A new search for time-reversal symmetry violation in the
  205{Tl} nucleus},\ }\href@noop {} {\bibfield  {journal} {\bibinfo  {journal}
  {Quantum Science and Technology}\ }\textbf {\bibinfo {volume} {6}},\ \bibinfo
  {pages} {044007} (\bibinfo {year} {2021})}\BibitemShut {NoStop}%
\bibitem [{\citenamefont {Fleig}\ and\ \citenamefont
  {DeMille}(2021)}]{DeMille2021}%
  \BibitemOpen
  \bibfield  {author} {\bibinfo {author} {\bibfnamefont {T.}~\bibnamefont
  {Fleig}}\ and\ \bibinfo {author} {\bibfnamefont {D.}~\bibnamefont
  {DeMille}},\ }\bibfield  {title} {\bibinfo {title} {Theoretical aspects of
  radium-containing molecules amenable to assembly from laser-cooled atoms for
  new physics searches},\ }\href {https://doi.org/10.1088/1367-2630/ac3619}
  {\bibfield  {journal} {\bibinfo  {journal} {New Journal of Physics}\ }\textbf
  {\bibinfo {volume} {23}},\ \bibinfo {pages} {113039} (\bibinfo {year}
  {2021})}\BibitemShut {NoStop}%
\bibitem [{\citenamefont {Singh}(2019)}]{Singh2019}%
  \BibitemOpen
  \bibfield  {author} {\bibinfo {author} {\bibfnamefont {J.~T.}\ \bibnamefont
  {Singh}},\ }\bibfield  {title} {\bibinfo {title} {A new concept for searching
  for time-reversal symmetry violation using {Pa}-229 ions trapped in optical
  crystals},\ }\bibfield  {journal} {\bibinfo  {journal} {Hyperfine
  Interactions}\ }\textbf {\bibinfo {volume} {240}},\ \href
  {https://doi.org/10.1007/s10751-019-1573-z} {10.1007/s10751-019-1573-z}
  (\bibinfo {year} {2019})\BibitemShut {NoStop}%
\bibitem [{\citenamefont {Flambaum}\ and\ \citenamefont
  {Dzuba}(2020)}]{Flambaum2020b}%
  \BibitemOpen
  \bibfield  {author} {\bibinfo {author} {\bibfnamefont {V.~V.}\ \bibnamefont
  {Flambaum}}\ and\ \bibinfo {author} {\bibfnamefont {V.~A.}\ \bibnamefont
  {Dzuba}},\ }\bibfield  {title} {\bibinfo {title} {{Electric dipole moments of
  atoms and molecules produced by enhanced nuclear Schiff moments}},\ }\href
  {https://doi.org/10.1103/PhysRevA.101.042504} {\bibfield  {journal} {\bibinfo
   {journal} {Phys. Rev. A}\ }\textbf {\bibinfo {volume} {101}},\ \bibinfo
  {pages} {1} (\bibinfo {year} {2020})}\BibitemShut {NoStop}%
\bibitem [{\citenamefont {Graner}\ \emph {et~al.}(2016)\citenamefont {Graner},
  \citenamefont {Chen}, \citenamefont {Lindahl},\ and\ \citenamefont
  {Heckel}}]{Graner2016}%
  \BibitemOpen
  \bibfield  {author} {\bibinfo {author} {\bibfnamefont {B.}~\bibnamefont
  {Graner}}, \bibinfo {author} {\bibfnamefont {Y.}~\bibnamefont {Chen}},
  \bibinfo {author} {\bibfnamefont {E.~G.}\ \bibnamefont {Lindahl}},\ and\
  \bibinfo {author} {\bibfnamefont {B.~R.}\ \bibnamefont {Heckel}},\ }\bibfield
   {title} {\bibinfo {title} {Reduced limit on the permanent electric dipole
  moment of $^{199}\mathrm{Hg}$},\ }\href
  {https://doi.org/10.1103/PhysRevLett.116.161601} {\bibfield  {journal}
  {\bibinfo  {journal} {Phys. Rev. Lett.}\ }\textbf {\bibinfo {volume} {116}},\
  \bibinfo {pages} {161601} (\bibinfo {year} {2016})}\BibitemShut {NoStop}%
\bibitem [{\citenamefont {Ramachandran}\ and\ \citenamefont
  {Vutha}(2022)}]{Ramachandran2022}%
  \BibitemOpen
  \bibfield  {author} {\bibinfo {author} {\bibfnamefont {H.~D.}\ \bibnamefont
  {Ramachandran}}\ and\ \bibinfo {author} {\bibfnamefont {A.~C.}\ \bibnamefont
  {Vutha}},\ }\bibfield  {title} {\bibinfo {title} {Nuclear {T}-violation
  search using octupolar nuclei in a crystal},\ }in\ \href@noop {} {\emph
  {\bibinfo {booktitle} {The 27th International Conference on Atomic
  Physics}}}\ (\bibinfo {year} {2022})\BibitemShut {NoStop}%
\bibitem [{\citenamefont {Flambaum}\ and\ \citenamefont
  {Feldmeier}(2020)}]{Flambaum2020}%
  \BibitemOpen
  \bibfield  {author} {\bibinfo {author} {\bibfnamefont {V.~V.}\ \bibnamefont
  {Flambaum}}\ and\ \bibinfo {author} {\bibfnamefont {H.}~\bibnamefont
  {Feldmeier}},\ }\bibfield  {title} {\bibinfo {title} {{Enhanced nuclear
  Schiff moment in stable and metastable nuclei}},\ }\href
  {https://doi.org/10.1103/PhysRevC.101.015502} {\bibfield  {journal} {\bibinfo
   {journal} {Phys. Rev. C}\ }\textbf {\bibinfo {volume} {101}},\ \bibinfo
  {pages} {1} (\bibinfo {year} {2020})}\BibitemShut {NoStop}%
\bibitem [{\citenamefont {Mirzai}\ \emph {et~al.}(2021)\citenamefont {Mirzai},
  \citenamefont {Ahadi}, \citenamefont {Melin},\ and\ \citenamefont
  {Olsson}}]{Mirzai2021}%
  \BibitemOpen
  \bibfield  {author} {\bibinfo {author} {\bibfnamefont {A.}~\bibnamefont
  {Mirzai}}, \bibinfo {author} {\bibfnamefont {A.}~\bibnamefont {Ahadi}},
  \bibinfo {author} {\bibfnamefont {S.}~\bibnamefont {Melin}},\ and\ \bibinfo
  {author} {\bibfnamefont {P.}~\bibnamefont {Olsson}},\ }\bibfield  {title}
  {\bibinfo {title} {First-principle investigation of doping effects on
  mechanical and thermodynamic properties of {Y2SiO5}},\ }\href
  {https://doi.org/https://doi.org/10.1016/j.mechmat.2020.103739} {\bibfield
  {journal} {\bibinfo  {journal} {Mechanics of Materials}\ }\textbf {\bibinfo
  {volume} {154}},\ \bibinfo {pages} {103739} (\bibinfo {year}
  {2021})}\BibitemShut {NoStop}%
\bibitem [{\citenamefont {Ferrier}\ \emph {et~al.}(2016)\citenamefont
  {Ferrier}, \citenamefont {Tumino},\ and\ \citenamefont
  {Goldner}}]{Ferrier2016}%
  \BibitemOpen
  \bibfield  {author} {\bibinfo {author} {\bibfnamefont {A.}~\bibnamefont
  {Ferrier}}, \bibinfo {author} {\bibfnamefont {B.}~\bibnamefont {Tumino}},\
  and\ \bibinfo {author} {\bibfnamefont {P.}~\bibnamefont {Goldner}},\
  }\bibfield  {title} {\bibinfo {title} {Variations in the oscillator strength
  of the 7 {F} 0 → 5 {D} 0 transition in {Eu} 3 + : {Y} 2 {SiO} 5 single
  crystals},\ }\href {https://doi.org/10.1016/j.jlumin.2015.07.026} {\bibfield
  {journal} {\bibinfo  {journal} {J. Luminescence}\ }\textbf {\bibinfo {volume}
  {170}},\ \bibinfo {pages} {406} (\bibinfo {year} {2016})}\BibitemShut
  {NoStop}%
\bibitem [{\citenamefont {Zhong}\ \emph {et~al.}(2015)\citenamefont {Zhong},
  \citenamefont {Hedges}, \citenamefont {Ahlefeldt}, \citenamefont
  {Bartholomew}, \citenamefont {Beavan}, \citenamefont {Wittig}, \citenamefont
  {Longdell},\ and\ \citenamefont {Sellars}}]{Sellars2015}%
  \BibitemOpen
  \bibfield  {author} {\bibinfo {author} {\bibfnamefont {M.}~\bibnamefont
  {Zhong}}, \bibinfo {author} {\bibfnamefont {M.~P.}\ \bibnamefont {Hedges}},
  \bibinfo {author} {\bibfnamefont {R.~L.}\ \bibnamefont {Ahlefeldt}}, \bibinfo
  {author} {\bibfnamefont {J.~G.}\ \bibnamefont {Bartholomew}}, \bibinfo
  {author} {\bibfnamefont {S.~E.}\ \bibnamefont {Beavan}}, \bibinfo {author}
  {\bibfnamefont {S.~M.}\ \bibnamefont {Wittig}}, \bibinfo {author}
  {\bibfnamefont {J.~J.}\ \bibnamefont {Longdell}},\ and\ \bibinfo {author}
  {\bibfnamefont {M.~J.}\ \bibnamefont {Sellars}},\ }\bibfield  {title}
  {\bibinfo {title} {Optically addressable nuclear spins in a solid with a
  six-hour coherence time},\ }\href {https://doi.org/10.1038/nature14025}
  {\bibfield  {journal} {\bibinfo  {journal} {Nature}\ }\textbf {\bibinfo
  {volume} {517}},\ \bibinfo {pages} {177} (\bibinfo {year}
  {2015})}\BibitemShut {NoStop}%
\bibitem [{\citenamefont {Timoney}\ \emph {et~al.}(2012)\citenamefont
  {Timoney}, \citenamefont {Lauritzen}, \citenamefont {Usmani}, \citenamefont
  {Afzelius},\ and\ \citenamefont {Gisin}}]{Timoney2012}%
  \BibitemOpen
  \bibfield  {author} {\bibinfo {author} {\bibfnamefont {N.}~\bibnamefont
  {Timoney}}, \bibinfo {author} {\bibfnamefont {B.}~\bibnamefont {Lauritzen}},
  \bibinfo {author} {\bibfnamefont {I.}~\bibnamefont {Usmani}}, \bibinfo
  {author} {\bibfnamefont {M.}~\bibnamefont {Afzelius}},\ and\ \bibinfo
  {author} {\bibfnamefont {N.}~\bibnamefont {Gisin}},\ }\bibfield  {title}
  {\bibinfo {title} {Atomic frequency comb memory with spin-wave storage in
  {$^{153}$Eu$^{3+}$:Y$_2$SiO$_5$}},\ }\href
  {https://doi.org/10.1088/0953-4075/45/12/124001} {\bibfield  {journal}
  {\bibinfo  {journal} {Journal of Physics B: Atomic, Molecular and Optical
  Physics}\ }\textbf {\bibinfo {volume} {45}},\ \bibinfo {pages} {124001}
  (\bibinfo {year} {2012})}\BibitemShut {NoStop}%
\bibitem [{\citenamefont {Zhong}\ and\ \citenamefont
  {Goldner}(2019)}]{Zhong2019}%
  \BibitemOpen
  \bibfield  {author} {\bibinfo {author} {\bibfnamefont {T.}~\bibnamefont
  {Zhong}}\ and\ \bibinfo {author} {\bibfnamefont {P.}~\bibnamefont
  {Goldner}},\ }\bibfield  {title} {\bibinfo {title} {Emerging rare-earth doped
  material platforms for quantum nanophotonics},\ }\href
  {https://doi.org/doi:10.1515/nanoph-2019-0185} {\bibfield  {journal}
  {\bibinfo  {journal} {Nanophotonics}\ }\textbf {\bibinfo {volume} {8}},\
  \bibinfo {pages} {2003} (\bibinfo {year} {2019})}\BibitemShut {NoStop}%
\bibitem [{\citenamefont {K\"onz}\ \emph {et~al.}(2003)\citenamefont {K\"onz},
  \citenamefont {Sun}, \citenamefont {Thiel}, \citenamefont {Cone},
  \citenamefont {Equall}, \citenamefont {Hutcheson},\ and\ \citenamefont
  {Macfarlane}}]{Konz2003}%
  \BibitemOpen
  \bibfield  {author} {\bibinfo {author} {\bibfnamefont {F.}~\bibnamefont
  {K\"onz}}, \bibinfo {author} {\bibfnamefont {Y.}~\bibnamefont {Sun}},
  \bibinfo {author} {\bibfnamefont {C.~W.}\ \bibnamefont {Thiel}}, \bibinfo
  {author} {\bibfnamefont {R.~L.}\ \bibnamefont {Cone}}, \bibinfo {author}
  {\bibfnamefont {R.~W.}\ \bibnamefont {Equall}}, \bibinfo {author}
  {\bibfnamefont {R.~L.}\ \bibnamefont {Hutcheson}},\ and\ \bibinfo {author}
  {\bibfnamefont {R.~M.}\ \bibnamefont {Macfarlane}},\ }\bibfield  {title}
  {\bibinfo {title} {Temperature and concentration dependence of optical
  dephasing, spectral-hole lifetime, and anisotropic absorption in
  ${\mathrm{eu}}^{3+}{:\mathrm{Y}}_{2}{\mathrm{sio}}_{5}$},\ }\href
  {https://doi.org/10.1103/PhysRevB.68.085109} {\bibfield  {journal} {\bibinfo
  {journal} {Phys. Rev. B}\ }\textbf {\bibinfo {volume} {68}},\ \bibinfo
  {pages} {085109} (\bibinfo {year} {2003})}\BibitemShut {NoStop}%
\bibitem [{\citenamefont {Cruzeiro}\ \emph {et~al.}(2018)\citenamefont
  {Cruzeiro}, \citenamefont {Etesse}, \citenamefont {Tiranov}, \citenamefont
  {Bourdel}, \citenamefont {Fr\"owis}, \citenamefont {Goldner}, \citenamefont
  {Gisin},\ and\ \citenamefont {Afzelius}}]{Cruzeiro2018}%
  \BibitemOpen
  \bibfield  {author} {\bibinfo {author} {\bibfnamefont {E.~Z.}\ \bibnamefont
  {Cruzeiro}}, \bibinfo {author} {\bibfnamefont {J.}~\bibnamefont {Etesse}},
  \bibinfo {author} {\bibfnamefont {A.}~\bibnamefont {Tiranov}}, \bibinfo
  {author} {\bibfnamefont {P.-A.}\ \bibnamefont {Bourdel}}, \bibinfo {author}
  {\bibfnamefont {F.}~\bibnamefont {Fr\"owis}}, \bibinfo {author}
  {\bibfnamefont {P.}~\bibnamefont {Goldner}}, \bibinfo {author} {\bibfnamefont
  {N.}~\bibnamefont {Gisin}},\ and\ \bibinfo {author} {\bibfnamefont
  {M.}~\bibnamefont {Afzelius}},\ }\bibfield  {title} {\bibinfo {title}
  {Characterization of the hyperfine interaction of the excited
  $^{5}\mathrm{D}_{0}$ state of
  {${\mathrm{Eu}}^{3+}:{\mathrm{Y}}_{2}{\mathrm{SiO}}_{5}$}},\ }\href
  {https://doi.org/10.1103/PhysRevB.97.094416} {\bibfield  {journal} {\bibinfo
  {journal} {Phys. Rev. B}\ }\textbf {\bibinfo {volume} {97}},\ \bibinfo
  {pages} {094416} (\bibinfo {year} {2018})}\BibitemShut {NoStop}%
\bibitem [{\citenamefont {Lauritzen}\ \emph {et~al.}(2012)\citenamefont
  {Lauritzen}, \citenamefont {Timoney}, \citenamefont {Gisin}, \citenamefont
  {Afzelius}, \citenamefont {de~Riedmatten}, \citenamefont {Sun}, \citenamefont
  {Macfarlane},\ and\ \citenamefont {Cone}}]{Afzelius2012}%
  \BibitemOpen
  \bibfield  {author} {\bibinfo {author} {\bibfnamefont {B.}~\bibnamefont
  {Lauritzen}}, \bibinfo {author} {\bibfnamefont {N.}~\bibnamefont {Timoney}},
  \bibinfo {author} {\bibfnamefont {N.}~\bibnamefont {Gisin}}, \bibinfo
  {author} {\bibfnamefont {M.}~\bibnamefont {Afzelius}}, \bibinfo {author}
  {\bibfnamefont {H.}~\bibnamefont {de~Riedmatten}}, \bibinfo {author}
  {\bibfnamefont {Y.}~\bibnamefont {Sun}}, \bibinfo {author} {\bibfnamefont
  {R.~M.}\ \bibnamefont {Macfarlane}},\ and\ \bibinfo {author} {\bibfnamefont
  {R.~L.}\ \bibnamefont {Cone}},\ }\bibfield  {title} {\bibinfo {title}
  {Spectroscopic investigations of {Eu${}^{3+}$:Y${}_{2}$SiO${}_{5}$} for
  quantum memory applications},\ }\href
  {https://doi.org/10.1103/PhysRevB.85.115111} {\bibfield  {journal} {\bibinfo
  {journal} {Phys. Rev. B}\ }\textbf {\bibinfo {volume} {85}},\ \bibinfo
  {pages} {115111} (\bibinfo {year} {2012})}\BibitemShut {NoStop}%
\bibitem [{\citenamefont {Arcangeli}\ \emph {et~al.}(2015)\citenamefont
  {Arcangeli}, \citenamefont {Macfarlane}, \citenamefont {Ferrier},\ and\
  \citenamefont {Goldner}}]{Arcangeli2015}%
  \BibitemOpen
  \bibfield  {author} {\bibinfo {author} {\bibfnamefont {A.}~\bibnamefont
  {Arcangeli}}, \bibinfo {author} {\bibfnamefont {R.~M.}\ \bibnamefont
  {Macfarlane}}, \bibinfo {author} {\bibfnamefont {A.}~\bibnamefont
  {Ferrier}},\ and\ \bibinfo {author} {\bibfnamefont {P.}~\bibnamefont
  {Goldner}},\ }\bibfield  {title} {\bibinfo {title} {Temperature dependence of
  nuclear spin coherence in
  {$\mathrm{E}{\mathrm{u}}^{3+}:{\mathrm{Y}}_{2}\mathrm{Si}{\mathrm{O}}_{5}$}},\
  }\href {https://doi.org/10.1103/PhysRevB.92.224401} {\bibfield  {journal}
  {\bibinfo  {journal} {Phys. Rev. B}\ }\textbf {\bibinfo {volume} {92}},\
  \bibinfo {pages} {224401} (\bibinfo {year} {2015})}\BibitemShut {NoStop}%
\bibitem [{\citenamefont {Zhang}\ \emph {et~al.}(2020)\citenamefont {Zhang},
  \citenamefont {Lučić}, \citenamefont {Galland}, \citenamefont {Le~Targat},
  \citenamefont {Goldner}, \citenamefont {Fang}, \citenamefont {Seidelin},\
  and\ \citenamefont {Le~Coq}}]{Zhang2020}%
  \BibitemOpen
  \bibfield  {author} {\bibinfo {author} {\bibfnamefont {S.}~\bibnamefont
  {Zhang}}, \bibinfo {author} {\bibfnamefont {N.}~\bibnamefont {Lučić}},
  \bibinfo {author} {\bibfnamefont {N.}~\bibnamefont {Galland}}, \bibinfo
  {author} {\bibfnamefont {R.}~\bibnamefont {Le~Targat}}, \bibinfo {author}
  {\bibfnamefont {P.}~\bibnamefont {Goldner}}, \bibinfo {author} {\bibfnamefont
  {B.}~\bibnamefont {Fang}}, \bibinfo {author} {\bibfnamefont {S.}~\bibnamefont
  {Seidelin}},\ and\ \bibinfo {author} {\bibfnamefont {Y.}~\bibnamefont
  {Le~Coq}},\ }\bibfield  {title} {\bibinfo {title} {Precision measurements of
  electric-field-induced frequency displacements of an ultranarrow optical
  transition in ions in a solid},\ }\href {https://doi.org/10.1063/5.0025356}
  {\bibfield  {journal} {\bibinfo  {journal} {Appl. Phys. Lett.}\ }\textbf
  {\bibinfo {volume} {117}},\ \bibinfo {pages} {221102} (\bibinfo {year}
  {2020})}\BibitemShut {NoStop}%
\bibitem [{\citenamefont {Kaplyanskii}\ and\ \citenamefont
  {McFarlane}(2012)}]{Kaplyanskii2012}%
  \BibitemOpen
  \bibfield  {author} {\bibinfo {author} {\bibfnamefont {A.}~\bibnamefont
  {Kaplyanskii}}\ and\ \bibinfo {author} {\bibfnamefont {R.}~\bibnamefont
  {McFarlane}},\ }\href@noop {} {\emph {\bibinfo {title} {Spectroscopy of
  Crystals Containing Rare Earth Ions}}}\ (\bibinfo  {publisher} {Elsevier},\
  \bibinfo {year} {2012})\BibitemShut {NoStop}%
\bibitem [{\citenamefont {Overhauser}\ and\ \citenamefont
  {R\"uchardt}(1958)}]{Overhauser1958}%
  \BibitemOpen
  \bibfield  {author} {\bibinfo {author} {\bibfnamefont {A.~W.}\ \bibnamefont
  {Overhauser}}\ and\ \bibinfo {author} {\bibfnamefont {H.}~\bibnamefont
  {R\"uchardt}},\ }\bibfield  {title} {\bibinfo {title} {Inversion symmetry of
  $m$ and $r$ centers},\ }\href@noop {} {\bibfield  {journal} {\bibinfo
  {journal} {Phys. Rev.}\ }\textbf {\bibinfo {volume} {112}},\ \bibinfo {pages}
  {722} (\bibinfo {year} {1958})}\BibitemShut {NoStop}%
\bibitem [{\citenamefont {Kaiser}\ \emph {et~al.}(1961)\citenamefont {Kaiser},
  \citenamefont {Sugano},\ and\ \citenamefont {Wood}}]{Kaiser1961}%
  \BibitemOpen
  \bibfield  {author} {\bibinfo {author} {\bibfnamefont {W.}~\bibnamefont
  {Kaiser}}, \bibinfo {author} {\bibfnamefont {S.}~\bibnamefont {Sugano}},\
  and\ \bibinfo {author} {\bibfnamefont {D.~L.}\ \bibnamefont {Wood}},\
  }\bibfield  {title} {\bibinfo {title} {Splitting of the emission lines of
  ruby by an external electric field},\ }\href@noop {} {\bibfield  {journal}
  {\bibinfo  {journal} {Phys. Rev. Lett.}\ }\textbf {\bibinfo {volume} {6}},\
  \bibinfo {pages} {605} (\bibinfo {year} {1961})}\BibitemShut {NoStop}%
\bibitem [{\citenamefont {Kaplyanskii}(2002)}]{Kaplyanskii2002}%
  \BibitemOpen
  \bibfield  {author} {\bibinfo {author} {\bibfnamefont {A.}~\bibnamefont
  {Kaplyanskii}},\ }\bibfield  {title} {\bibinfo {title} {Linear stark effect
  in spectroscopy and luminescence of doped inorganic insulating crystals},\
  }\href@noop {} {\bibfield  {journal} {\bibinfo  {journal} {J. Luminescence}\
  }\textbf {\bibinfo {volume} {100}},\ \bibinfo {pages} {21} (\bibinfo {year}
  {2002})}\BibitemShut {NoStop}%
\bibitem [{\citenamefont {Smith}\ \emph {et~al.}(2022)\citenamefont {Smith},
  \citenamefont {Reid}, \citenamefont {Sellars},\ and\ \citenamefont
  {Ahlefeldt}}]{Smith2022}%
  \BibitemOpen
  \bibfield  {author} {\bibinfo {author} {\bibfnamefont {K.~M.}\ \bibnamefont
  {Smith}}, \bibinfo {author} {\bibfnamefont {M.~F.}\ \bibnamefont {Reid}},
  \bibinfo {author} {\bibfnamefont {M.~J.}\ \bibnamefont {Sellars}},\ and\
  \bibinfo {author} {\bibfnamefont {R.~L.}\ \bibnamefont {Ahlefeldt}},\
  }\bibfield  {title} {\bibinfo {title} {Complete crystal-field calculation of
  zeeman hyperfine splittings in europium},\ }\href
  {https://doi.org/10.1103/PhysRevB.105.125141} {\bibfield  {journal} {\bibinfo
   {journal} {Phys. Rev. B}\ }\textbf {\bibinfo {volume} {105}},\ \bibinfo
  {pages} {125141} (\bibinfo {year} {2022})}\BibitemShut {NoStop}%
\bibitem [{\citenamefont {Longdell}\ \emph {et~al.}(2006)\citenamefont
  {Longdell}, \citenamefont {Alexander},\ and\ \citenamefont
  {Sellars}}]{Longdell2006}%
  \BibitemOpen
  \bibfield  {author} {\bibinfo {author} {\bibfnamefont {J.~J.}\ \bibnamefont
  {Longdell}}, \bibinfo {author} {\bibfnamefont {A.~L.}\ \bibnamefont
  {Alexander}},\ and\ \bibinfo {author} {\bibfnamefont {M.~J.}\ \bibnamefont
  {Sellars}},\ }\bibfield  {title} {\bibinfo {title} {Characterization of the
  hyperfine interaction in europium-doped yttrium orthosilicate and europium
  chloride hexahydrate},\ }\href {https://doi.org/10.1103/PhysRevB.74.195101}
  {\bibfield  {journal} {\bibinfo  {journal} {Phys. Rev. B}\ }\textbf {\bibinfo
  {volume} {74}},\ \bibinfo {pages} {195101} (\bibinfo {year}
  {2006})}\BibitemShut {NoStop}%
\bibitem [{\citenamefont {Stone}(2005)}]{Stone2005}%
  \BibitemOpen
  \bibfield  {author} {\bibinfo {author} {\bibfnamefont {N.}~\bibnamefont
  {Stone}},\ }\bibfield  {title} {\bibinfo {title} {Table of nuclear magnetic
  dipole and electric quadrupole moments},\ }\href
  {https://doi.org/https://doi.org/10.1016/j.adt.2005.04.001} {\bibfield
  {journal} {\bibinfo  {journal} {Atomic Data and Nuclear Data Tables}\
  }\textbf {\bibinfo {volume} {90}},\ \bibinfo {pages} {75} (\bibinfo {year}
  {2005})}\BibitemShut {NoStop}%
\bibitem [{\citenamefont {Klein}(1952)}]{Klein1952}%
  \BibitemOpen
  \bibfield  {author} {\bibinfo {author} {\bibfnamefont {M.~J.}\ \bibnamefont
  {Klein}},\ }\bibfield  {title} {\bibinfo {title} {On a degeneracy theorem of
  {K}ramers},\ }\href@noop {} {\bibfield  {journal} {\bibinfo  {journal} {Am.
  J. Phys.}\ }\textbf {\bibinfo {volume} {20}},\ \bibinfo {pages} {65}
  (\bibinfo {year} {1952})}\BibitemShut {NoStop}%
\bibitem [{\citenamefont {Yano}\ \emph {et~al.}(1991)\citenamefont {Yano},
  \citenamefont {Mitsunaga},\ and\ \citenamefont {Uesugi}}]{Yano1991}%
  \BibitemOpen
  \bibfield  {author} {\bibinfo {author} {\bibfnamefont {R.}~\bibnamefont
  {Yano}}, \bibinfo {author} {\bibfnamefont {M.}~\bibnamefont {Mitsunaga}},\
  and\ \bibinfo {author} {\bibfnamefont {N.}~\bibnamefont {Uesugi}},\
  }\bibfield  {title} {\bibinfo {title} {Ultralong optical dephasing time in
  {Eu3$+$:Y2SiO5}},\ }\href {https://doi.org/10.1364/OL.16.001884} {\bibfield
  {journal} {\bibinfo  {journal} {Opt. Lett.}\ }\textbf {\bibinfo {volume}
  {16}},\ \bibinfo {pages} {1884} (\bibinfo {year} {1991})}\BibitemShut
  {NoStop}%
\bibitem [{\citenamefont {Yano}\ \emph {et~al.}(1992)\citenamefont {Yano},
  \citenamefont {Mitsunaga},\ and\ \citenamefont {Uesugi}}]{Yano1992}%
  \BibitemOpen
  \bibfield  {author} {\bibinfo {author} {\bibfnamefont {R.}~\bibnamefont
  {Yano}}, \bibinfo {author} {\bibfnamefont {M.}~\bibnamefont {Mitsunaga}},\
  and\ \bibinfo {author} {\bibfnamefont {N.}~\bibnamefont {Uesugi}},\
  }\bibfield  {title} {\bibinfo {title} {Nonlinear laser spectroscopy of
  {Eu3$+$:Y2SiO5} and its application to time-domain optical memory},\ }\href
  {https://doi.org/10.1364/JOSAB.9.000992} {\bibfield  {journal} {\bibinfo
  {journal} {J. Opt. Soc. Am. B}\ }\textbf {\bibinfo {volume} {9}},\ \bibinfo
  {pages} {992} (\bibinfo {year} {1992})}\BibitemShut {NoStop}%
\bibitem [{\citenamefont {Macfarlane}\ and\ \citenamefont
  {Shelby}(1987)}]{Macfarlane1987}%
  \BibitemOpen
  \bibfield  {author} {\bibinfo {author} {\bibfnamefont {R.}~\bibnamefont
  {Macfarlane}}\ and\ \bibinfo {author} {\bibfnamefont {R.}~\bibnamefont
  {Shelby}},\ }\bibfield  {title} {\bibinfo {title} {Coherent transient and
  holeburning spectroscopy of rare earth ions in solids},\ }in\ \href@noop {}
  {\emph {\bibinfo {booktitle} {Modern Problems in Condensed Matter
  Sciences}}},\ Vol.~\bibinfo {volume} {21}\ (\bibinfo  {publisher}
  {Elsevier},\ \bibinfo {year} {1987})\ pp.\ \bibinfo {pages}
  {51--184}\BibitemShut {NoStop}%
\bibitem [{\citenamefont {Macfarlane}\ \emph {et~al.}(2014)\citenamefont
  {Macfarlane}, \citenamefont {Arcangeli}, \citenamefont {Ferrier},\ and\
  \citenamefont {Goldner}}]{Macfarlane2014}%
  \BibitemOpen
  \bibfield  {author} {\bibinfo {author} {\bibfnamefont {R.}~\bibnamefont
  {Macfarlane}}, \bibinfo {author} {\bibfnamefont {A.}~\bibnamefont
  {Arcangeli}}, \bibinfo {author} {\bibfnamefont {A.}~\bibnamefont {Ferrier}},\
  and\ \bibinfo {author} {\bibfnamefont {P.}~\bibnamefont {Goldner}},\
  }\bibfield  {title} {\bibinfo {title} {Optical measurement of the effect of
  electric fields on the nuclear spin coherence of rare-earth ions in solids},\
  }\href {https://doi.org/10.1103/PhysRevLett.113.157603} {\bibfield  {journal}
  {\bibinfo  {journal} {Phys. Rev. Lett.}\ }\textbf {\bibinfo {volume} {113}},\
  \bibinfo {pages} {157603} (\bibinfo {year} {2014})}\BibitemShut {NoStop}%
\bibitem [{\citenamefont {Arcangeli}\ \emph {et~al.}(2014)\citenamefont
  {Arcangeli}, \citenamefont {Lovri\ifmmode~\acute{c}\else \'{c}\fi{}},
  \citenamefont {Tumino}, \citenamefont {Ferrier},\ and\ \citenamefont
  {Goldner}}]{Arcangeli2014}%
  \BibitemOpen
  \bibfield  {author} {\bibinfo {author} {\bibfnamefont {A.}~\bibnamefont
  {Arcangeli}}, \bibinfo {author} {\bibfnamefont {M.}~\bibnamefont
  {Lovri\ifmmode~\acute{c}\else \'{c}\fi{}}}, \bibinfo {author} {\bibfnamefont
  {B.}~\bibnamefont {Tumino}}, \bibinfo {author} {\bibfnamefont
  {A.}~\bibnamefont {Ferrier}},\ and\ \bibinfo {author} {\bibfnamefont
  {P.}~\bibnamefont {Goldner}},\ }\bibfield  {title} {\bibinfo {title}
  {Spectroscopy and coherence lifetime extension of hyperfine transitions in
  {${}^{151}\mathrm{Eu}{}^{3+}{\mathrm{:Y}}_{2}{\mathrm{SiO}}_{5}$}},\ }\href
  {https://doi.org/10.1103/PhysRevB.89.184305} {\bibfield  {journal} {\bibinfo
  {journal} {Phys. Rev. B}\ }\textbf {\bibinfo {volume} {89}},\ \bibinfo
  {pages} {184305} (\bibinfo {year} {2014})}\BibitemShut {NoStop}%
\bibitem [{\citenamefont {Ramachandran}\ \emph {et~al.}(2023)\citenamefont
  {Ramachandran}, \citenamefont {Ford},\ and\ \citenamefont
  {Vutha}}]{Ramachandran2023}%
  \BibitemOpen
  \bibfield  {author} {\bibinfo {author} {\bibfnamefont {H.~D.}\ \bibnamefont
  {Ramachandran}}, \bibinfo {author} {\bibfnamefont {J.~E.}\ \bibnamefont
  {Ford}},\ and\ \bibinfo {author} {\bibfnamefont {A.~C.}\ \bibnamefont
  {Vutha}},\ }\href@noop {} {\bibinfo {title} {Coherent quantum beats:
  spectroscopy of energy differences buried within inhomogeneous broadening}}
  (\bibinfo {year} {2023}),\ \Eprint {https://arxiv.org/abs/2304.06189}
  {arXiv:2304.06189 [physics.atom-ph]} \BibitemShut {NoStop}%
\bibitem [{\citenamefont {DeMille}\ \emph {et~al.}(2000)\citenamefont
  {DeMille}, \citenamefont {Bay}, \citenamefont {Bickman}, \citenamefont
  {Kawall}, \citenamefont {Krause~Jr}, \citenamefont {Maxwell},\ and\
  \citenamefont {Hunter}}]{DeMille2000}%
  \BibitemOpen
  \bibfield  {author} {\bibinfo {author} {\bibfnamefont {D.}~\bibnamefont
  {DeMille}}, \bibinfo {author} {\bibfnamefont {F.}~\bibnamefont {Bay}},
  \bibinfo {author} {\bibfnamefont {S.}~\bibnamefont {Bickman}}, \bibinfo
  {author} {\bibfnamefont {D.}~\bibnamefont {Kawall}}, \bibinfo {author}
  {\bibfnamefont {D.}~\bibnamefont {Krause~Jr}}, \bibinfo {author}
  {\bibfnamefont {S.}~\bibnamefont {Maxwell}},\ and\ \bibinfo {author}
  {\bibfnamefont {L.}~\bibnamefont {Hunter}},\ }\bibfield  {title} {\bibinfo
  {title} {Investigation of pbo as a system for measuring the electric dipole
  moment of the electron},\ }\href@noop {} {\bibfield  {journal} {\bibinfo
  {journal} {Phys. Rev. A}\ }\textbf {\bibinfo {volume} {61}},\ \bibinfo
  {pages} {052507} (\bibinfo {year} {2000})}\BibitemShut {NoStop}%
\bibitem [{\citenamefont {C.~Carvalho}\ \emph {et~al.}(2015)\citenamefont
  {C.~Carvalho}, \citenamefont {Le~Floch}, \citenamefont {Krupka},\ and\
  \citenamefont {Tobar}}]{Carvalho2015}%
  \BibitemOpen
  \bibfield  {author} {\bibinfo {author} {\bibfnamefont {N.}~\bibnamefont
  {C.~Carvalho}}, \bibinfo {author} {\bibfnamefont {J.-M.}\ \bibnamefont
  {Le~Floch}}, \bibinfo {author} {\bibfnamefont {J.}~\bibnamefont {Krupka}},\
  and\ \bibinfo {author} {\bibfnamefont {M.}~\bibnamefont {Tobar}},\ }\bibfield
   {title} {\bibinfo {title} {Multi-mode technique for the determination of the
  biaxial y2sio5 permittivity tensor from 300 to 6 k},\ }\href
  {https://doi.org/10.1063/1.4920987} {\bibfield  {journal} {\bibinfo
  {journal} {Applied Physics Letters}\ }\textbf {\bibinfo {volume} {106}}
  (\bibinfo {year} {2015})}\BibitemShut {NoStop}%
\bibitem [{\citenamefont {Clavaguéra}\ and\ \citenamefont
  {Dognon}(2005)}]{Clavaguera2005}%
  \BibitemOpen
  \bibfield  {author} {\bibinfo {author} {\bibfnamefont {C.}~\bibnamefont
  {Clavaguéra}}\ and\ \bibinfo {author} {\bibfnamefont {J.}~\bibnamefont
  {Dognon}},\ }\bibfield  {title} {\bibinfo {title} {Accurate static electric
  dipole polarizability calculations of +3 charged lanthanide ions},\ }\href
  {https://doi.org/10.1016/j.chemphys.2004.10.014} {\bibfield  {journal}
  {\bibinfo  {journal} {Chem. Phys.}\ }\textbf {\bibinfo {volume} {311}},\
  \bibinfo {pages} {169} (\bibinfo {year} {2005})}\BibitemShut {NoStop}%
\bibitem [{\citenamefont {Nishimura}\ \emph {et~al.}(1991)\citenamefont
  {Nishimura}, \citenamefont {Tanaka}, \citenamefont {Kurita},\ and\
  \citenamefont {Kushida}}]{Nishimura1991}%
  \BibitemOpen
  \bibfield  {author} {\bibinfo {author} {\bibfnamefont {G.}~\bibnamefont
  {Nishimura}}, \bibinfo {author} {\bibfnamefont {M.}~\bibnamefont {Tanaka}},
  \bibinfo {author} {\bibfnamefont {A.}~\bibnamefont {Kurita}},\ and\ \bibinfo
  {author} {\bibfnamefont {T.}~\bibnamefont {Kushida}},\ }\bibfield  {title}
  {\bibinfo {title} {{5D0}-{7F0} transition mechanism of {Eu3}+ in {Ca}({PO3})2
  glass, {Y2O2S} crystal and polyvinyl alcohol},\ }\href
  {https://doi.org/10.1016/0022-2313(91)90172-R} {\bibfield  {journal}
  {\bibinfo  {journal} {J. Luminescence}\ }\textbf {\bibinfo {volume} {48}},\
  \bibinfo {pages} {473} (\bibinfo {year} {1991})}\BibitemShut {NoStop}%
\bibitem [{Note1()}]{Note1}%
  \BibitemOpen
  \bibinfo {note} {A calculation for a similar system, EuCl$_3$, has been
  presented in Ref. \cite {Sushkov2023}.}\BibitemShut {Stop}%
\bibitem [{\citenamefont {Abel}\ \emph {et~al.}(2020)\citenamefont {Abel} \emph
  {et~al.}}]{Abel2020}%
  \BibitemOpen
  \bibfield  {author} {\bibinfo {author} {\bibfnamefont {C.}~\bibnamefont
  {Abel}} \emph {et~al.},\ }\bibfield  {title} {\bibinfo {title} {Measurement
  of the permanent electric dipole moment of the neutron},\ }\href
  {https://doi.org/10.1103/PhysRevLett.124.081803} {\bibfield  {journal}
  {\bibinfo  {journal} {Phys. Rev. Lett.}\ }\textbf {\bibinfo {volume} {124}},\
  \bibinfo {pages} {081803} (\bibinfo {year} {2020})}\BibitemShut {NoStop}%
\bibitem [{\citenamefont {Graham}\ and\ \citenamefont
  {Rajendran}(2013)}]{Graham2013}%
  \BibitemOpen
  \bibfield  {author} {\bibinfo {author} {\bibfnamefont {P.~W.}\ \bibnamefont
  {Graham}}\ and\ \bibinfo {author} {\bibfnamefont {S.}~\bibnamefont
  {Rajendran}},\ }\bibfield  {title} {\bibinfo {title} {New observables for
  direct detection of axion dark matter},\ }\href
  {https://doi.org/10.1103/PhysRevD.88.035023} {\bibfield  {journal} {\bibinfo
  {journal} {Phys. Rev. D}\ }\textbf {\bibinfo {volume} {88}},\ \bibinfo
  {pages} {035023} (\bibinfo {year} {2013})}\BibitemShut {NoStop}%
\bibitem [{\citenamefont {Roussy}\ \emph {et~al.}(2021)\citenamefont {Roussy}
  \emph {et~al.}}]{Roussy2021}%
  \BibitemOpen
  \bibfield  {author} {\bibinfo {author} {\bibfnamefont {T.~S.}\ \bibnamefont
  {Roussy}} \emph {et~al.},\ }\bibfield  {title} {\bibinfo {title}
  {Experimental constraint on axionlike particles over seven orders of
  magnitude in mass},\ }\href@noop {} {\bibfield  {journal} {\bibinfo
  {journal} {Phys. Rev. Lett.}\ }\textbf {\bibinfo {volume} {126}},\ \bibinfo
  {pages} {171301} (\bibinfo {year} {2021})}\BibitemShut {NoStop}%
\bibitem [{\citenamefont {Aybas}\ \emph {et~al.}(2021)\citenamefont {Aybas}
  \emph {et~al.}}]{Aybas2021}%
  \BibitemOpen
  \bibfield  {author} {\bibinfo {author} {\bibfnamefont {D.}~\bibnamefont
  {Aybas}} \emph {et~al.},\ }\bibfield  {title} {\bibinfo {title} {Search for
  axionlike dark matter using solid-state nuclear magnetic resonance},\ }\href
  {https://doi.org/10.1103/PhysRevLett.126.141802} {\bibfield  {journal}
  {\bibinfo  {journal} {Phys. Rev. Lett.}\ }\textbf {\bibinfo {volume} {126}},\
  \bibinfo {pages} {141802} (\bibinfo {year} {2021})}\BibitemShut {NoStop}%
\bibitem [{\citenamefont {Sushkov}\ \emph {et~al.}(2023)\citenamefont
  {Sushkov}, \citenamefont {Sushkov},\ and\ \citenamefont
  {Yaresko}}]{Sushkov2023}%
  \BibitemOpen
  \bibfield  {author} {\bibinfo {author} {\bibfnamefont {A.~O.}\ \bibnamefont
  {Sushkov}}, \bibinfo {author} {\bibfnamefont {O.~P.}\ \bibnamefont
  {Sushkov}},\ and\ \bibinfo {author} {\bibfnamefont {A.}~\bibnamefont
  {Yaresko}},\ }\href@noop {} {\bibinfo {title} {Effective electric field:
  quantifying the sensitivity of searches for new {P,T}-odd physics with
  {EuCl$_3\cdot$6H$_2$O}}} (\bibinfo {year} {2023}),\ \Eprint
  {https://arxiv.org/abs/2304.08461} {arXiv:2304.08461 [physics.atom-ph]}
  \BibitemShut {NoStop}%
\end{thebibliography}%

\newpage
\appendix
\onecolumngrid

\section{Coherent quantum beats}\label{sec:cqb}
\begin{figure}
  \centering
  \includegraphics[width=0.4\textwidth]{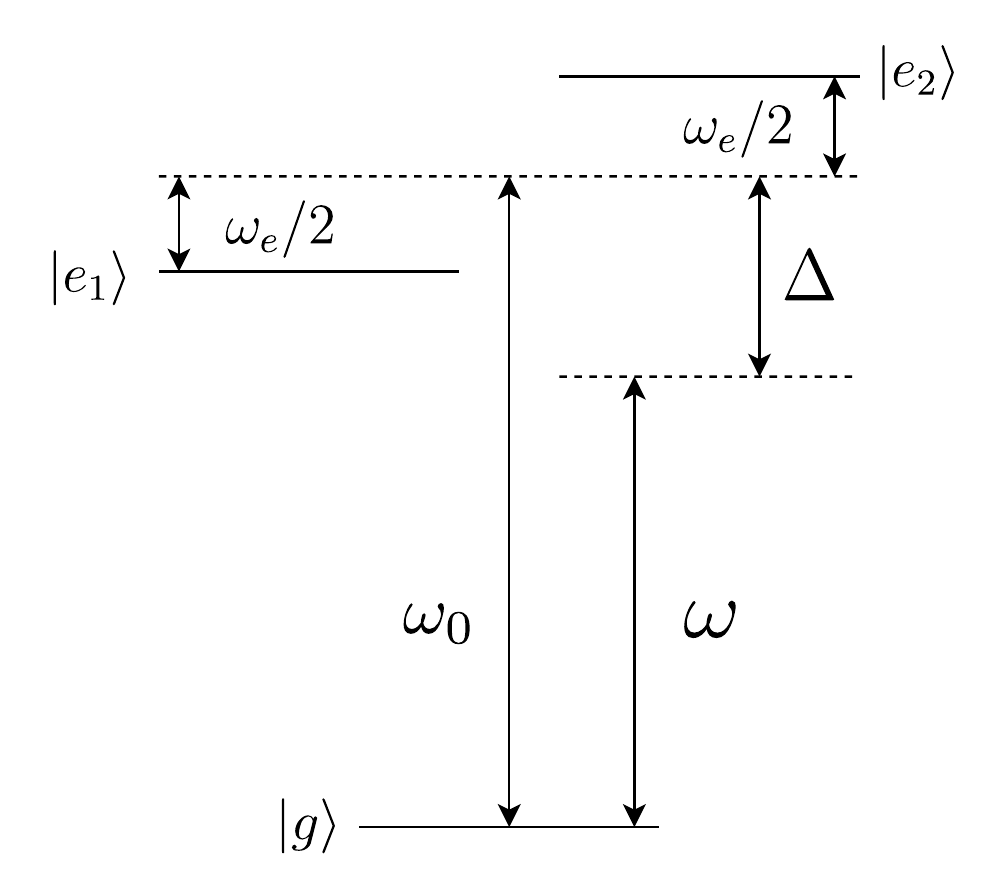}
  \caption{A system of three levels, where the energy difference of interest ($\omega_e$) is measured using spectroscopy of the $g \to e_1$ and $g \to e_2$ transitions. The frequency of light is $\omega$ and the average of the $g \to e_1,e_2$ resonance frequencies is $\omega_0$. The driving frequency is detuned from the resonant frequency by $\Delta$. This figure is reproduced from Ref.\ \cite{Ramachandran2023}.}
  \label{fig:three_level}
\end{figure}

The coherent quantum beat (CQB) method allows for the measurement of a small frequency difference $\omega_e$ masked by the much larger inhomogeneous width $\Gamma_\mrm{inh}$. We summarize the main features of CQB with a simple three-level system, shown in Fig.\ \ref{fig:three_level}. Consider a pair of near-degenerate states $e_1, e_2$, separated by $\omega_e$, and a third state $g$, separated by $\omega_0$ from the center of $e_1, e_2$. The atom interacts with light at the frequency $\omega$, detuned by an amount $\Delta = \omega - \omega_0$, which drives the inhomogeneously-broadened $g \to e_1, e_2$ transitions.

The CQB measurement sequence consists of two $\pi$-pulses separated by a free evolution time $T$. When $g \to e_1, e_2$ is driven, interference occurs between $e_1, e_2$ and persists even when $\Gamma_\mrm{inh} \gg \omega_e$. To see this, note that the detuning $\Delta$ affects the amplitudes $\alpha$ and $\beta$ in the coherent superposition $\ket{\psi(0)} = \alpha \ket{e_1} + \beta \ket{e_2}$ prepared by the first pulse, but it does not affect the phase accumulated during the free evolution to the state $\ket{\psi(T)} = \alpha \ket{e_1} e^{+i \omega_e T/2}+ \beta \ket{e_2} e^{-i \omega_e T/2}$. This final state $\ket{\psi(T)}$ is projected onto $g$ by the second pulse. The CQB sequence leads to a modulation in the ground-state population $P_g$  as a function of $T$, at the frequency $\omega_e$. Further, the modulation is  \textbf{$\Delta$-insensitive}, allowing for a precise measurement of $\omega_e$ even when it is much smaller than $\Gamma_\mathrm{inh}$. 

In Ref.\ \cite{Ramachandran2023}, this idea is developed analytically and numerically, and extended to four-level systems consisting of two pairs of near-degenerate states $g_1, g_2$ and $e_1, e_2$ in initial mixed states. In particular, the CQB method can be applied to the four-level system $\{a,\bar{a},b,\bar{b}\}$ (described in Section \ref{sec:scheme}) to measure the desired energy difference $\nu_{\bar{a}a}$. Higher precision can be obtained compared to conventional spectroscopy techniques such as Ramsey interferometry, since all the atoms in the inhomogeneously-broadened ensemble contribute to the measurement.

\section{Matrix elements of the magnetic dipole moment operator}\label{sec:matrix_elements}
This section summarizes the matrix elements of the magnetic dipole moment operator between hyperfine states in the $^{7}$F$_0$ electronic state of $^{153}$Eu:YSO in zero magnetic field, along the three crystallographic axes of the host crystal. These values were calculated using the constants measured in Refs.\ \cite{Yano1991, Longdell2006}, adjusted as described in Section \ref{sec:structure}.

\begin{table}[h!]
    \centering
    
    \begin{tabular}{lrrrrrr}
        \toprule
        {} &  $\ket{a}$ &  $\ket{\bar{a}}$ &  $\ket{b}$ &  $\ket{\bar{b}}$ &  $\ket{c}$ &  $\ket{\bar{c}}$ \\
        \midrule
        $\bra{a}$       &      0.746 &            0.021 &      0.218 &            0.109 &      0.054 &            0.050 \\
        $\bra{\bar{a}}$ &      0.021 &            0.746 &      0.109 &            0.218 &      0.050 &            0.054 \\
        $\bra{b}$       &      0.218 &            0.109 &      0.351 &            0.220 &      0.299 &            0.167 \\
        $\bra{\bar{b}}$ &      0.109 &            0.218 &      0.220 &            0.351 &      0.167 &            0.299 \\
        $\bra{c}$       &      0.054 &            0.050 &      0.299 &            0.167 &      0.025 &            0.110 \\
        $\bra{\bar{c}}$ &      0.050 &            0.054 &      0.167 &            0.299 &      0.110 &            0.025 \\
        \bottomrule
    \end{tabular}
    \caption{Matrix element magnitudes $|\bra{j}\vec{\mu}\ket{i}|$ of the magnetic dipole operator $\vec{\mu} = \mathbf{M} \cdot \vec{I}$ along the $D1$ axis in units of the nuclear magneton $\mu_n$.}
    \label{tab:D1_matrix_elements}
\end{table}

\begin{table}[h!]
    \centering
    
    \begin{tabular}{lrrrrrr}
        \toprule
        {} &  $\ket{a}$ &  $\ket{\bar{a}}$ &  $\ket{b}$ &  $\ket{\bar{b}}$ &  $\ket{c}$ &  $\ket{\bar{c}}$ \\
        \midrule
        $\bra{a}$       &      0.487 &            0.020 &      0.231 &            0.040 &      0.056 &            0.027 \\
        $\bra{\bar{a}}$ &      0.020 &            0.487 &      0.040 &            0.231 &      0.027 &            0.056 \\
        $\bra{b}$       &      0.231 &            0.040 &      0.248 &            0.198 &      0.247 &            0.039 \\
        $\bra{\bar{b}}$ &      0.040 &            0.231 &      0.198 &            0.248 &      0.039 &            0.247 \\
        $\bra{c}$       &      0.056 &            0.027 &      0.247 &            0.039 &      0.436 &            0.091 \\
        $\bra{\bar{c}}$ &      0.027 &            0.056 &      0.039 &            0.247 &      0.091 &            0.436 \\
        \bottomrule
    \end{tabular}
    \caption{Matrix element magnitudes $|\bra{j}\vec{\mu}\ket{i}|$ of the magnetic dipole operator $\vec{\mu} = \mathbf{M} \cdot \vec{I}$ along the $D2$ axis in units of $\mu_n$.}
    \label{tab:D2_matrix_elements}
\end{table}

\begin{table}[h!]
    \centering
    
    \begin{tabular}{lrrrrrr}
        \toprule
        {} &  $\ket{a}$ &  $\ket{\bar{a}}$ &  $\ket{b}$ &  $\ket{\bar{b}}$ &  $\ket{c}$ &  $\ket{\bar{c}}$ \\
        \midrule
        $\bra{a}$       &      0.199 &            0.040 &      0.568 &            0.188 &      0.098 &            0.089 \\
        $\bra{\bar{a}}$ &      0.040 &            0.199 &      0.188 &            0.568 &      0.089 &            0.098 \\
        $\bra{b}$       &      0.568 &            0.188 &      0.402 &            0.367 &      0.403 &            0.354 \\
        $\bra{\bar{b}}$ &      0.188 &            0.568 &      0.367 &            0.402 &      0.354 &            0.403 \\
        $\bra{c}$       &      0.098 &            0.089 &      0.403 &            0.354 &      1.169 &            0.123 \\
        $\bra{\bar{c}}$ &      0.089 &            0.098 &      0.354 &            0.403 &      0.123 &            1.169 \\
        \bottomrule
    \end{tabular}
    \caption{Matrix element magnitudes $|\bra{j}\vec{\mu}\ket{i}|$ of the magnetic dipole operator $\vec{\mu} = \mathbf{M} \cdot \vec{I}$ along the $b$ axis in units of $\mu_n$.}
    \label{tab:b_matrix_elements}
\end{table}

\end{document}